\documentclass[10pt,journal,compsoc]{IEEEtran}
\usepackage{graphicx}
\usepackage{amsmath}
\usepackage{enumerate}
\usepackage{multirow}
\usepackage{array}
\usepackage{float}
\usepackage{subfigure}
\usepackage{algorithm}
\usepackage{algorithmicx}
\usepackage{algpseudocode}
\usepackage{epstopdf}
\usepackage{balance}
\balance

\begin{document}
%
\title{PIS: A Multi-dimensional Routing Protocol for Socially-aware Networking}

\author{Feng Xia,~\IEEEmembership{Senior Member, IEEE}, Li Liu, Behrouz Jedari, and Sajal K. Das,~\IEEEmembership{Fellow, IEEE} 
\thanks{F. Xia, L. Liu and B. Jedari are with the School of Software, Dalian University of Technology, China.}
\thanks{S. K. Das is with the Computer Science Department, Missouri University of Science and Technology, USA.}
\thanks{Corresponding author: Feng Xia; E-mail: f.xia@ieee.org.}
\thanks{Manuscript received XX XX, 2014; revised XX XX, 20XX.}
}

\markboth{IEEE Transactions on Mobile Computing, Vol. X, No. X, XX 20XX}%
{Liu \MakeLowercase{\textit{et al.}}: PIS: A Multi-dimensional Routing Protocol in Socially-aware Networking}

\IEEEcompsoctitleabstractindextext{%
\begin{abstract}
Socially-aware networking is an emerging paradigm for intermittently connected networks consisting of mobile users with social relationships and characteristics. In this setting, humans are the main carriers of mobile devices.
Hence, their connections, social features, and behaviors can be exploited to improve the performance of data forwarding protocols. In this paper, we first explore the impact of three social features, namely physical proximity, user interests, and social relationship on users' daily routines. Then, we propose a multi-dimensional routing protocol called Proximity-Interest-Social (PIS) protocol in which the three different social dimensions are integrated into a unified distance function in order to select optimal intermediate data carriers. PIS protocol utilizes a time slot management mechanism to discover users' movement similarities in different time periods during a day. We compare the performance of PIS to Epidemic, PROPHET, and SimBet routing protocols using SIGCOMM09 and INFOCOM06 data sets. The experiment results show that PIS outperforms other benchmark routing protocols with the highest data delivery ratio with a low communication overhead.

\end{abstract}

}

\maketitle

\IEEEdisplaynotcompsoctitleabstractindextext

%
\IEEEpeerreviewmaketitle

\section{Introduction}
\IEEEPARstart{T}{oday} a large volume of mobile data traffic is transferred via 3G cellular networks which make them overloaded. To tackle this problem, proximity-based wireless technologies such as Bluetooth and Wi-Fi have been used as promising solutions to lighten the network traffic among mobile users. Socially-aware networking (SAN) \cite{Xia2013Socially} is a new trend of delay tolerant networks (DTNs) \cite{DTNs} and opportunistic communications \cite{Oppnet}\cite{Isento2013Vehicular}, which exploits social properties of mobile carriers (i.e., users) to design efficient networking protocols. In this setting, mobile nodes employ \textit{store-carry-and-forward} mechanism to communicate with each other with the help of their short-distance and low-cost devices in order to share data objects (e.g., pictures, advertisements, software updates) among interested users. The basic idea of SAN is to exploit the users' social context information to improve the performance of data routing protocols. To achieve this goal, spatio-temporal and connectivity properties of the mobile users are captured in an efficient way.

When a mobile node contacts others, all the encountered nodes could be candidates for relaying her messages to the destination node. Nevertheless, selecting appropriate encounter nodes is a very challenging problem. One potential approach is to forward a message to nodes which are closer to the destination (a greedy strategy). In this context, some investigations on human movement patterns have been carried out to explore the spatio-temporal properties of mobile users and predict their future contacts \cite{MobilitySurvey2014, MobPre1, MobPre2}. Since the physical location of mobile users vary, more stable social factors should be considered to make efficient and effective routing decisions. To overcome the shortcomings in mobility based routing approaches, several social-aware routing protocols for SAN paradigm have been proposed in the literature (see \cite{Routing-survey} for a survey). Some recent social-aware routing protocols such as \cite{Bubble-rap, Zhu2014SMART, HSFR, CAOR} make forwarding decisions based on the users' social characteristics and similarities. In particular, social network analysis techniques are used to extract meaningful social relationships among mobile users. The main motivation is that the social behaviors of mobile nodes have long-term characteristics, which provide reliable connectivity among them.

In this paper, we take multiple social characteristics of users into consideration to design a stable routing protocol in the SAN paradigm. Considering the users' daily routines, we observe that three social factors- namely physical proximity, interests, and social relationship- influence the performance of data forwarding. The \textit{physical proximity} indicates direct contacts among mobile nodes. When two nodes come into the communication range of each other, they can exchange their messages. The physical proximity is a popular factor which is used in many existing routing protocols extensively \cite{Bubble-rap, Routing:Daly, Information:Daly}. The second factor is the user \textit{interests} which indicate their preferences for data. Based on Homophily concept \cite{Homophily}, individuals with common interests tend to meet each other more often and perform similar actions. Furthermore, their interests are stable for a long time. The third factor is \textit{social relationships} such as friendship, family, or colleague which describes the users' personal relations. Nodes with strong social relationships contact each other more frequently and regularly. Friendship based routing \cite{Bulut-Friendship} is a well-known example in this regard which identify the social ties among mobile nodes based on their contact history and thus form social communities. The basic idea of our novel approach lies in the fact that integration of these social attributes can be utilized to improve the overall efficiency of a routing protocol since each of these parameters can affect a data delivery protocol in different time periods. For example, One (or more) factors may work in the current time period while others may work in a future time period. It may also be possible that these three factors work simultaneously in some other occasions.

We provide an example to illustrate our idea. Let us consider Alice's regular mobility pattern in different time periods during a working day. She takes a bus from 7:30 to 8:30, works between 8:30 and 17:00, goes to a gym to exercise from 17:00 to 19:00, and comes back home at the end of the day. In each time period, different social characteristics are exhibited. Moreover, she could share data with different people such as strangers in a bus (geographic proximity), colleagues in her office (colleague relationship), friends (common interests) and family members (family relationship). It is difficult to determine which social characteristic always plays the decisive role throughout the day. Therefore, Alice could select appropriate forwarders by comprehensively fusing social properties of people around her, as well as temporal dimensions (time regularity).
\begin{figure}[!t]
\centering
\includegraphics[width=3 in]{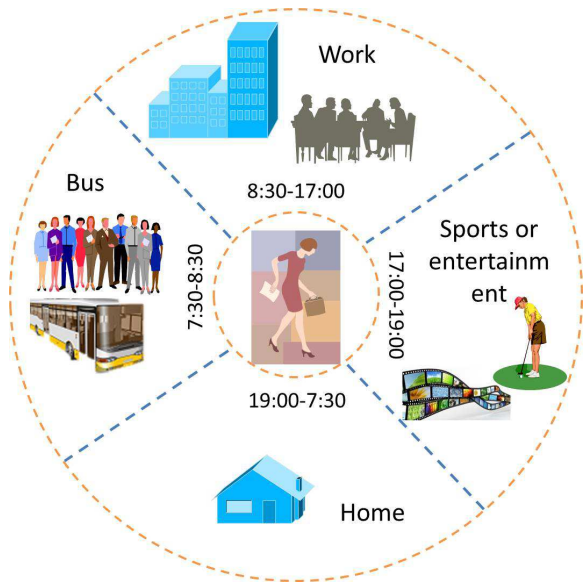}
\caption{Human mobility: regularity and social characteristics in daily routine.}
\label{socialproperties}
\end{figure}

In this paper, we propose a multi-dimensional routing protocol, namely Proximity-Interest-Social relationship (PIS) protocol for socially-aware networking (SAN) paradigm. The PIS protocol integrates multiple social dimensions of mobile users in a utility function to select the best intermediate nodes and improve the overall routing performance. In this method, a time slot management mechanism is used to manage the social information of users in different time periods.

Our major contributions can be summarized as follows:
\begin{itemize}
\item We analyze real mobility traces of mobile users and explore three social factors: physical proximity, user interests, and social relationships in order to make a stable and adaptable routing protocol in the SAN paradigm. Then, we introduce a multi-dimensional routing protocol, called PIS protocol based on the social properties of mobile nodes.
\item Based on the users' daily routine, we present a time slot management mechanism in PIS to keeptrack of contact records, self/contact interests, and direnct/indirect social relationship information. This mechanism uses the social properties of mobile users in different time slots to reflect the  movement routines in different time periods.
\item We apply an efficient copy control mechanism to control data congestion in the PIS protocol and decrease the network overhead.
\item We compare PIS to three benchmark routing protocols, Epidemic \cite{Epidemic}, PROPHET \cite{PROPHET}, and SimBet \cite{Routing:Daly} using SIGCOMM09 \cite{sigcomm09} and INFOCOM06 \cite{scott2009crawdad} real world traces. The experimental results demonstrate that our protocol guarantee higher performance of routing which achieves the highest data delivery ratio with a low communication overhead.
\end{itemize}

The remainder of this paper is organized as follows. Section II provides a review of related works on SAN. Sections III and IV presents the PIS protocol and its implementation details. Section V describes the performance evaluation results of PIS in comparison with other well-known routing protocols. Finally, we conclude the paper in Section VI.

\section{Related Work}
It is hard to guarantee a stable end-to-end delivery path among mobile nodes in the SAN paradigm. Therefore, message delivery becomes a challenging issue in this setting. To Tackle this problem, several data routing protocols have been proposed in intermittently connected networks. A potential solution is data flooding in which the whole generated data are transferred to other encountered nodes. Flooding-based protocols such as Epidemic routing \cite{Epidemic} with an unlimited number of message replications waste the network resources dramatically and cause data congestion. To cope with these problems, several single-copy or multi-copy routing protocols have been proposed aiming to limit the number of message copies and leverage a tradeoff between resource usage and message delivery probability. As an improvement to the Epidemic routing, PROPHET \cite{PROPHET} routing uses a delivery predictability metric to calculate how likely a node will be able to deliver a message to its destination. In this method, the contact frequency between the nodes is used as a context information.

Recently, social aspects of mobile users have been widely utilized to streamline routing decisions in the SAN paradigm (see \cite{Routing-survey} for a survey). The main reason is that social attributes and relations of mobile users have generally long-term characteristics and they are less volatile. Essentially, these methods attempt to group nodes into communities and/or choose a node with high centrality or similarity (e.g., interests, context, common friends, etc.) with the destination node as the packet forwarder.

For example, Bubble Rap \cite{Bubble-rap} is a prominent social-based data forwarding algorithm which focuses on social centrality. The algorithm structures nodes into communities with different sizes based on social parameters. High popularity nodes and community members of the destination are selected as relays. Besides, nodes have various levels of popularity (i.e., rank). Each node has a global ranking and a local ranking. A node's global ranking denotes its popularity in the entire society, with its local ranking within its own community. Messages are forwarded according to the global ranking until a node in the destination community is found. Then, the messages are forwarded in the destination community via nodes with higher local ranking. Similarly, LocalCom \cite{LocalCom} forms the community structure and considers the forwarding between different communities. First, similarity metrics according to the nodes' contact history are calculated to construct the neighboring graph. Next, a distributed algorithm detects the underlying community structure. Finally, LocalCom selects and prunes gateways to connect communities, control redundancy and facilitate efficient inter-community forwarding.

Friendship describes close personal relationships as a concept in sociology. In SAN, the friendship can be defined between a pair of nodes. In other words, two nodes need to have long-lasting and regular contacts to be friends of one another. Friendship based routing \cite{Bulut-Friendship} considers  the contact history of mobile nodes to measure their social ties and form friendship community. This method considers three behavioral features of close friendship: high frequency, longevity, and regularity, and two metrics called social pressure metric (SPM) and conditional SPM are defined for direct and indirect friendship.

Based on the Homophily principle \cite{Homophily}, nodes with similar interests tend to meet each other more often. Common interests between users have led to many innovative protocols. As an example, social-aware networking (SANE) \cite{SANE} is a pioneering forwarding protocol extended from the Epidemic routing. In SANE, nodes exchange their interest profiles and then each node starts scanning its buffer for messages to relay. In \cite{SPOON}, a peer-to-peer content-based file sharing system, called SPOON, is proposed to take advantage of interest. SPOON extracts a node's interest from its files and groups common-interest nodes as communities. The authors in \cite{user-centric}  propose a user-centric data dissemination protocol in DTNs where the contact patterns and interests of mobile users are exploited to measure the nodes' centrality. Similarly, SocialCast \cite{P-Socially} is a data dissemination protocol based on publish/subscribe systems which considers the nodes' social ties and mobility patterns as well as their interests to select next intermediate nodes.

In addition to the above-mentioned protocols, BEEINFO \cite{XiaBEEINFO} is proposed as a combination of interest and swarm intelligence to choose the best forwarder. BEEINFO is inspired by artificial bee colony algorithm and adopts density and social tie to perform the inter-community and intra-community forwarding processes. Although the communities are classified by user interests, the exact classification is not described and further exploration on interest is not provided.

Some studies have also considered the user mobility regularities into consideration to improve the routing performance. As an example, Habit \cite{Habit} constructs the regularity graph and interest graph for routing protocol by leveraging information of nodes' movement regularity and their social networks. In HiBOp \cite{Hibop}, each node exchanges context information only with its community members and stores their context information. In the forwarding phase, only nodes of the destination community are selected as candidate forwarders. It selects packet forwarder according to the nodes' contact history. In fact, HiBOp looks for nodes which have the highest match with the known context attributes of the destination node. SEDUM \cite{SEDUM} is a social network oriented and duration utility-based routing protocol that considers movement patterns of mobile users to increase routing performance. The duration utility in this method is the ratio of total contact duration between two nodes over a time period. SEDUM considers both contact frequency and duration in movement patterns of network nodes. In addition, it uses multi-copy routing which can discover the minimum number of copies of a message using an optimal tree replication algorithm.

Some recent works combine multi social features to exploit valuable information. In \cite{zhou-spatial}, the spatial and temporal characteristics of mobile users are analyzed and temporal communities are constructed in order to predict their future contact probabilities. Similarly, CiPRO \cite{Nguyen-Context} considers both spatial and temporal dimensions of human mobility to predict the context of nodes so that the source device knows when and where to start the routing process to maximize the transmission delay and minimize the network overhead. M-Dimension routing \cite{L-multidimensional} exploits local information derived from multiple dimensions such as geographic and social dimensions. ML-SOR \cite{s-opportunistic} extracts social network information from multiple social contexts and combines three measures: node centrality, tie strength and a tie predictor thereby avoiding suboptimal paths. Finally, the authors in \cite{w-onexploiting} observed that the transient social contact patterns of mobile nodes during short time periods are usually different from their cumulative contact patterns. Then they exploited the transient social contact patterns from three aspects including transient contact distribution, transient connectivity, and transient community structure to improve the forwarding performance.

Considering the shortcomings of the above-mentioned methods, in this paper, we combine three social dimensions of mobile users in a utility function to select the best relay nodes in message forwarding. In our proposed multi-dimensional protocol, user movement during a day is divided into several time periods and a time slot management mechanism is used to measure the similarity of the users with respect to their social features. We present the details of the PIS protocol in the next Section.

\section{PIS Routing Protocol}

We propose a multi-dimensional routing protocol, called PIS protocol, which exploits the nodes' physical proximity, interests, and social relationship information to deliver messages among them in an adaptive and reliable way. Compared to existing social-based routing protocols, PIS mainly relies on multi-dimensional social properties and time regularities in order to improve the performance of data routing.

\begin{figure}[!t]
\centering
\includegraphics[width=3 in]{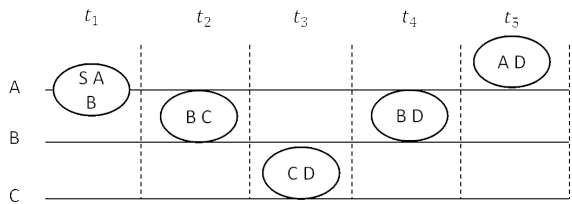}
\caption{An example of multi-dimensional relaying.}
\label{example}
\end{figure}

Considering the sample scenario as described in Section 1, the three social factors considered in PIS coexist and interweave in real SAN applications. Thus, they can effectively be used to identify appropriate intermediate nodes in different time periods of a routing process. To clarify the functionality of the PIS protocol, we provide a example routing in Fig. \ref{example}. In this figure, node S wants to deliver its messages to node D. Node A, B, and C are intermediate nodes. Nodes S, A, and B contact each other in \textit{t1}. Note that A and D have higher physical proximity in time period \textit{t5} while, C and D have the most common interests in \textit{t3}. Node B frequently contacts others which share their common interests with D in \textit{t2}. Thus, it has higher opportunity to meet nodes which have common interests with D (like node C). The friendship among nodes B and D has the highest value in \textit{t4}. Therefore, the path S-B-C-D is the best forwarding path between nodes S and D to relay their messages in the shortest time. Also, B will be the best forwarder. Integrating different social factors, the more appropriate forwarder will be selected to implement more stable delivery, even in different times or scenarios.

The physical proximity is the most reliable feature to design a routing protocol. In PIS, the physical proximity refers to the direct contacts among the nodes. We introduce ego network \cite{Information:Daly} to obtain the proximity information of the mobile nodes by utilizing their contact history. However, considering the limited power and storage resources in the mobile devices, they cannot maintain the global network state and can thus only capture the information of their contacting nodes (i.e., local information).

In order to overcome the restrictions of the physical location information, we consider user interests which are more stable and easy to maintain. We call the user interests as \textit{self-interest}, while there is another kind of interests called \textit{contact interest}. To explain the concept of the contact interest, we assume that there is a concentrated area of nodes within a common interest, such as football. The other nodes which are not interested in football join and leave this area frequently. Therefore, they have advantages to relay messages to this area. On the other hand, the users' self-interest information includes the topics they are interested in. In addition, we consider the users' social relationships in the PIS protocol. We consider both the \textit{direct} and \textit{indirect} social relationships, where the \textit{direct} relationships refers to their friends, while the \textit{indirect} relationship refers to the friends of their friends.

\begin{figure}[!t]
\centering
\includegraphics[width=3.5in]{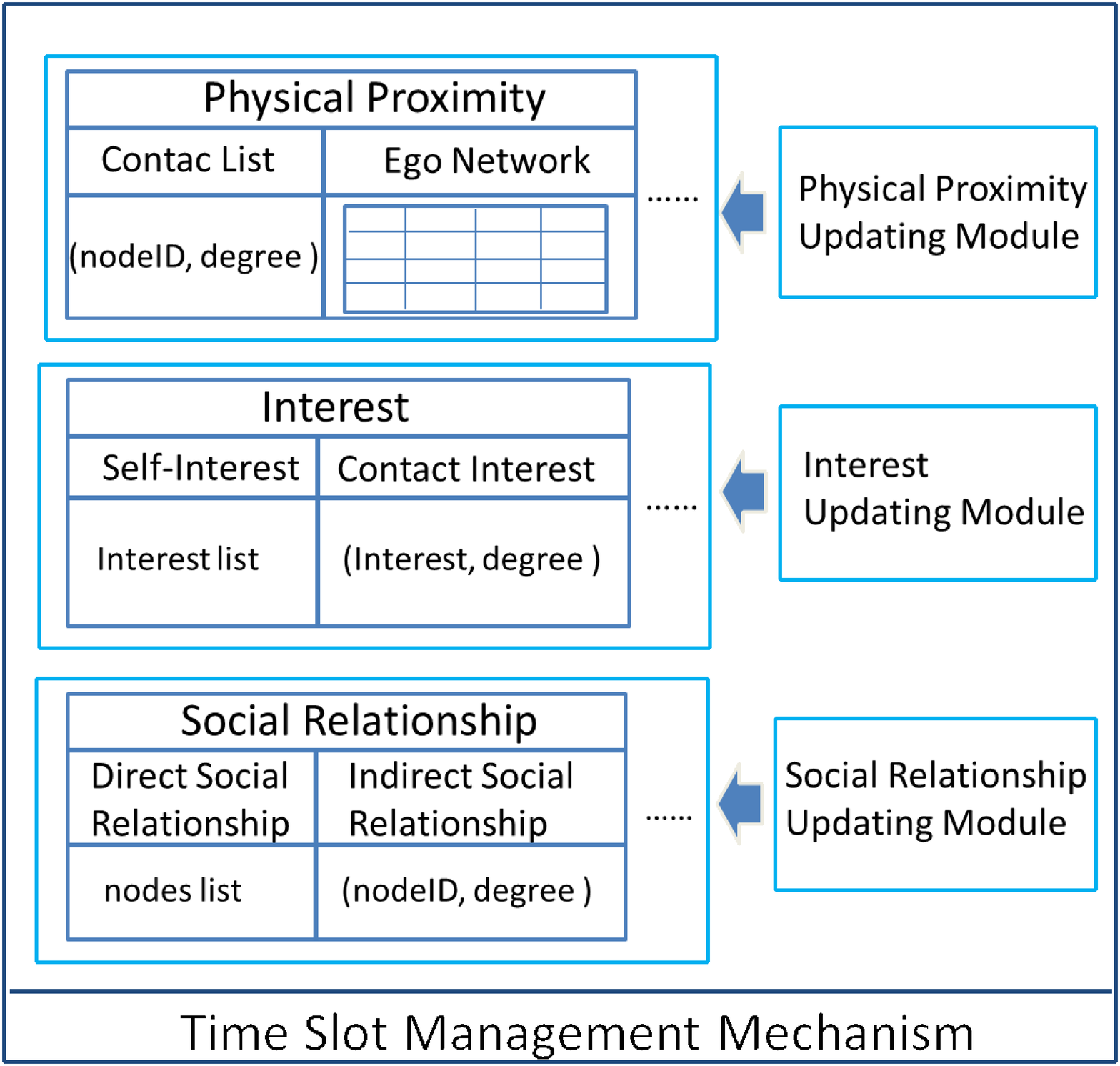}
\caption{Structure of a time slot in the time slot management mechanism.}
\label{archecture}
\end{figure}

Role of the three social features applied in the PIS protocol can be observed in the mobile users' daily routines on a regular basis. This information are used in the PIS protocol to predict the future contacts of mobile nodes. To this aim, a day is divided into several time slots where each social feature is managed in a distinct time slot. Then, we design a time slot management mechanism including distinct modules to process the information in each time slot. For instance, if a time slot includes two hours, a day is divided into 12 time slots. Fig. \ref{archecture} illustrates the structure of a time slot which is divided into three components: \textit{Physical Proximity}, \textit{Interest} and \textit{Social relationship}. Furthermore, three modules are designed to maintain the updated information in each time slot. We discuss this idea with more details in the following subsections.

\subsection{Physical Proximity}

The physical proximity includes the nodes' frequent and direct contact information in every time slot. Obviously, the location information of mobile nodes can be different since they may visit different places. Nevertheless, it is revealed that mobile users normally move between 4 and 6 major locations which occupy more than 70\% of their movements locations \cite{s-where}. Considering this idea, each mobile node in the PIS protocol maintains an \textit{ego network} in order to maintain the physical proximity of other nodes where a mobile node is considered as an "age". In other words, the \textit{ego network} is a matrix which reveals the contact information of other nodes. In the PIS protocol, an ego network is constructed by each node locally without having complete knowledge of the entire network. Then, in each time slot, a node records the frequency of its direct contact information. At the same time, it updates the ego network for each encountered node locally.

The structure of \textit{Physical Proximity} module consists of \textit{contact list} and \textit{ego network} as shown in Fig. \ref{archecture}. The \textit{contact list} includes a set of $(nodeID, degree)$ pairs, where \textit{nodeID} shows the identity of encountered nodes and \textit{degree} represents their respective contact frequency (encountered time) within a time slot. The \textit{Physical Proximity Updating Module} maintains and analyzes the contact and ego network information of the nodes during a particular time slot.

\subsection{User Interest}

The authors in \cite{a-creating} reveal that 80\% of computer files fall into 20\% of their total file categories. Taking this idea into consideration, 76 participants in SIGCOMM2009 data set \cite{sigcomm09} have 711 interest topics. However, the number of topics with at least two people interested in is 50, that is only 7\%. When we use applications or services installed in portable devices, our interest is easy to gather and analyze from our operations. Possibly, when we connect to mobile social networks, our interest list can be downloaded from the Internet directly. Additionally, the function of fringe nodes should not be ignored. For instance, there may be a nodes concentrated area for one interest. The fringe nodes are not interested in that interest, but frequently pass the area. Therefore, the fringe nodes are also the good forwarders. Considering utilizing fringe nodes effectively, the interest metric records not only the node's self-interest and degree, but also the encountered nodes' contact interest information (for fringe nodes).

The interest structure consists of two parts: \textit{Self-Interest} ($S_{I}$) and \textit{Contact Interest} ($C_{I}$). Both of them store a list of interests. $S_{I}$ records the nodes's self-interests. $C_{I}$, which consists of a list of $(interest, degree)$ pair, records interests of encountered nodes and their contact degrees. When a fringe node has a higher degree of one kind of interest, it will be higher opportunities to encounter the nodes with this interest. Then, it can be selected as a candidate forwarders. The \textit{Interest updating module} is responsible for recording and updating the information of \textit{Self-Interest} and \textit{Contact Interest}.

\subsection{Social Relationship}

\begin{figure}[!b]
\centering
\includegraphics[width=3in]{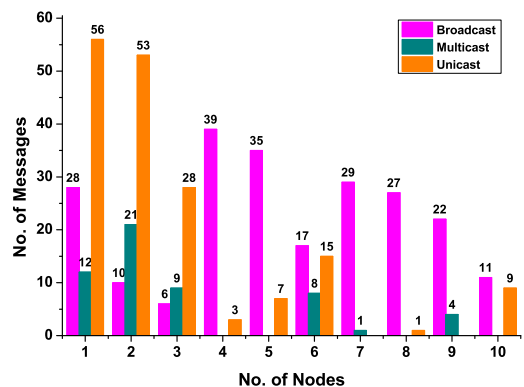}
\caption{The number of broadcast, multicast, and unicast messages in the SIGCOMM2009 data set.}
\label{message}
\end{figure}

The strength of the social relationships among mobile users can be used to predict their contact probabilities. In the SIGCOMM2009 data set, 898 messages are generated where the number of unicast messages is 263 (29.3\%). Since unicast happens between friends, top 10 nodes generated 457 messages and the number of unicast messages is 172 (37.6\%). Therefore, the friendship is an important feature to explore the number of message transmissions among source and destination nodes. In our analysis, we select top 10 nodes which generate the highest number of messages. Fig. \ref{message} shows the number of generated broadcast, multicast, and unicast messages in the SIGCOMM2009 data set which demonstrates that friendship is an important property in users' communication pattern

The PIS protocol considers both \textit{direct social relationships} ($D_{So}$) and \textit{indirect social relationships} ($In_{So}$). Taking the users' friendship information into consideration, the indirect social relationships indicate the friends of friends. When two users have common friends, they are expected to have higher probability to contact each other. $D_{So}$ denotes the set of nodes with direct social relationships whereas $In_{So}$ denotes the set of nodes with indirect social relationships. Let $In_{So}$ denotes the set of $(nodeID, degree)$ pairs which includes the strength of social relationships among mobile nodes. We design the \textit{social relationship updating module} to maintain and update the social relationship information.

\section{Implementation of PIS}

Mobile nodes in the SAN paradigm have homophily phenomenon which is the tendency of individuals to associate and bond with similar minded others. This is often expressed in the adage "birds of a feather flock together". The main idea of the PIS protocol relies on the homophily phenomenon from three social dimensions.

When two nodes A (also denoted $N_{A}$) and B (or $N_{B}$) contact each other, they update their contact list and exchange their interests, social information, and message list. For a sample message in the message list, PIS compares the similarities of nodes A and B to the destination node D with respect to the three dimensions. Then, it calculates the similarity utility function to select an appropriate forwarding node according to the similarities of the nodes.

\begin{figure}[!b]
\centering
\includegraphics[width=3in]{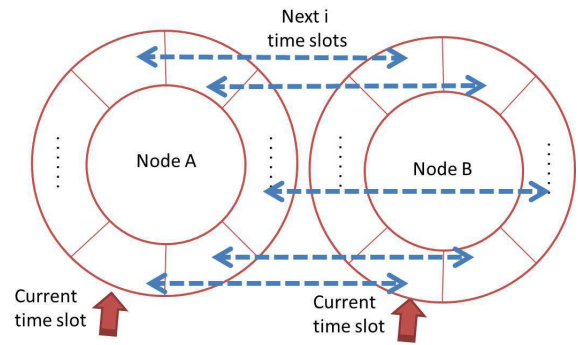}
\caption{Cycle and periodic time slots.}
\label{timeslot}
\end{figure}

Based on the time slot management mechanism, we get a cycle and periodic time slot, as shown in Fig. \ref{timeslot}. PIS manages the information of physical proximity, interest, and social relationship in different time slots. In order to make a forwarding decision, PIS compares the corresponding similarities in the last time slot $i$. The time slot $i$ indicates the current time while $i-1$ shows the next one. Generally, the parameter $i$ is closely related to the remaining TTL (time-to-live) of the messages. For example, if TTL of a message is 10 hours and time slot is 2 hours, $i$ can be assigned as 5 since the message will be dropped after 10 hours. PIS compares the similarities between the nodes in current and next 4 time slots, which represents the contact probabilities in 10 hours. In addition, we set different weights to each time slot such that the closer is the time slot to the current time, the larger is the weight.

Essential notations and symbols are summarized in Table \ref{notation}. The detailed message exchange process between nodes A and B is outlined in four steps as follows. Without loss of generality, we only describe the process from the viewpoint of source node A, while the same process happens to node B.

\begin{table}[!t]
\centering
\caption{Explanation of Notations}
\label{notation}
\begin{tabular}{c c}                                              \hline
Notation             & Explanation                        \\ \hline
A, B                 & Node A, Node B                        \\ 
D                    & Destination node                       \\ 
$S_{I}$              & Self-Interest                         \\ 
$C_{I}$              & Contact Interest                       \\ 
$D_{So}$             & Direct social relationship                      \\ 
$In_{So}$            & Indirect social relationship                       \\ 
$simPro_{A}$         & Proximity similarity between nodes A and D     \\ 
$simDevPro$          & Proximity similarity deviation        \\ 
$simIns_{A}$         & Interest similarity between nodes A and D   \\ 
$simDevIns$          & Interest similarity deviation           \\ 
$simSoc_{A}$         & social relationship similarity between nodes A and D \\ 
$simDevSoc$          & social relationship similarity deviation   \\ 
$simPIS$             & Similarity utility function                   \\ \hline
\end{tabular}
\end{table}

1) First, nodes A and B update their contact list and exchange their social information including \textit{contact list}, $S_{I}$ and $D_{So}$ in order to update the information of their ego network, i.e., $C_{I}$ and $In_{So}$.

2) Then, nodes A and B exchange their message list. For each message in the list of node B, PIS computes the similarities between nodes A and D with respect to physical proximity, interests, and social relationship, respectively, which are denoted as $simPro_{A}$, $simIns_{A}$ and $simSoc_{A}$.

3) The similarities between nodes B and D are calculated and attached in messages when they received, and denoted as $simPro_{B}$, $simIns_{B}$, and $simSoc_{B}$. Then, PIS computes the similarity deviation values among nodes A and B including $simDevPro$, $simDevIns$, and $simDevSoc$ .

4) Finally, the forwarding utility, denoted as $simPIS$, is calculated in order to make the forwarding decision. In the last step, PIS takes advantage of copy control mechanism in order to control data congestion in a data routing process.

\subsection{Information Update}

PIS utilizes the three social dimensions and time slot management mechanism to make a forwarding decision. When nodes contact, they exchange their contact list, $S_{I}$, and $D_{So}$ in order to update their ego network information, $C_{I}$, and $In_{So}$.

\subsubsection{Physical Proximity Update}

In the SAN paradigm, the topology of the network changes dynamically due to the mobility of nodes. Meanwhile, it is difficult for each node to maintain the information of all the other nodes. Ego network is a popular solution used in the intermittent connection networks such as SAN to tackle this issue. Nodes construct and maintain their own ego network information according to their contact information. The structure of an ego network in PIS is designed using a \textit{Ego} matrix as follows:

\begin{displaymath}
Ego=\left\{ \begin{array}{ll}
e_{A,B} & \textrm{if A contacts B}\\
0 & \textrm{otherwise}
\end{array} \right.
\end{displaymath}
where $e_{A,B}$ denotes the contact time among nodes A and B in a time slot. If nodes A and B contact each other during this period, $Ego_{A,B}$ equals $e_{A,B}$ which indicates the contact frequency of those nodes. Otherwise, $Ego_{A,B}$ is set to 0.
\begin{algorithm}[!b]
  \caption{Physical Proximity Updating}
  \label{proximity-update}
    \begin{algorithmic}[1]
    \State //In current time slot
    \State Update contact list
    \State //Update Ego of node A
    \State Exchange contact list
    \For {$N_{C}$ in the contact list of node B }
       \State //$N_{C}$ is the number of contacts of node B
       \State //degree is the contact frequency of nodes C and B
       \State $Ego_{C,B}$ = $degree$
       \State $Ego_{B,C}$ = $degree$
    \EndFor
  \end{algorithmic}
\end{algorithm}

The contact list of the nodes stores their identity as well as their contact time in a time slot. When two nodes A and B meet each other, they update their contact information and exchange their contact list in order to update the \textit{Ego} matrix. For example, node A receives the contact list of node B and updates the ego matrix of node A. Algorithm Physical Proximity Updating presents the updating process of the physical proximity information.

\subsubsection{Interest and Social Relationship Update}

Management of user interests and social relationships in PIS has similar procedure, and their updating strategies are also similar. Two encounter nodes A and B exchange values $S_{I}$ and $D_{So}$ in order to update $C_{I}$ and $In_{So}$. Note that $C_{I}$ and $In_{So}$ adopts similar structure $(interest/nodeID, degree)$, where \textit{degree} indicates the contact frequency with this interest/node. Therefore, we only describe the updating process of the user interests in Algorithm Interest Updating from node A's perspective.

For each interest $S_{I}$ of node B, the updating module checks whether node A includes $C_{I}$ or not. If $C_{I}$ contains the interest, it means that node A has met some nodes with the same interests as before. Therefore, her interest's degree is added by $incrementalValue$ to represent her meeting frequency. Otherwise, it implies that it is the first time that node A meets the interest. Next, node A adds the interest and its degree is initialized with $initialValue$.

\begin{algorithm}[!b]
  \caption{Interest Updating}
  \label{interest-update}
    \begin{algorithmic}[1]
    \State //In current time slot
    \State //$S_I$ is the self-interest of node B
    \State //$insB$  is the member of $S_I$
    \State //$C_I$ is the contact interest of node A
    \For {$insB$ in $S_I$ }
       \If {$C_I$.contains($insB$)}
            \State $C_I$.getValue($InsB$)+=$incrementalValue$
       \Else
            \State $C_I$.add($InsB$, $initialValue$)
       \EndIf
   \EndFor
  \end{algorithmic}
\end{algorithm}

\subsection{Social Similarity Measurement}

PIS compares the similarities between intermediate nodes and destination node on the three social features to choose an appropriate intermediate node. For the social features, PIS applies similar strategies to compute their similarities. Here, we compute the similarity of physical proximity ($simPro$), as an example, to describe the similarity measurement method. Parameter $SimPro_{A}$ indicates the physical proximity similarity between nodes A and D in \textit{i} time slots. The value of $simPro_{A}$ indicates the contact probability of the nodes.

In each time slot $t\in \{0, 1, ..., i-1\}$, the similarity between nodes A and D indicates the number of their common contact nodes, which is denoted as $simPro_{t}$. We aim to find an intermediate node which meets the destination node earlier. Therefore, we assign a parameter $\beta$ to adjust the weight of time slots ($\beta < 1$). The weight of time slot $t$ is denoted as $\beta^{t+1}$. In this way, the social properties in closer time slot will be highlighted and the optimal forwarding node will be selected. The value of $simPro$ is quantified by Equation \eqref{simPro}. The similarity computing algorithm of physical proximity is illustrated in Algorithm Proximity Similarity Calculation.

Both interest and social relationship similarities consist of two parts: $S_{I}$/$C_{I}$ and $D_{So}$/$In_{So}$. For the interest feature, the similarity of $S_I$ and $C_I$ are computed by Equation \eqref{simins} and the final $simIns$ can be obtained according to Equation \eqref{siminsf}. Similarly, Equations \eqref{simSoc} and \eqref{simSocf} describe how to calculate $simSoc$. In Equations \eqref{siminsf} and \eqref{simSocf}, $\alpha$ is a parameter to adjust the weight of two parts.

\begin{equation}\label{simPro}
 simPro = \sum_{t=0}^{i-1} simPro_{t} \times \beta^{t+1}
\end{equation}

\begin{equation}\label{simins}
 simIns\{S_I, C_I\} = \sum_{t=0}^{i-1} simIns_{t}\{S_I, C_I\} \times \beta^{t+1}
\end{equation}

\begin{equation}\label{siminsf}
 simIns = \alpha \times simIns\{S_I\} + (1-\alpha) \times simIns\{C_I\}
\end{equation}

\begin{equation}\label{simSoc}
simSoc\{D_{So}, In_{So}\} = \sum_{t=0}^{i-1} simSoc_{t}\{D_{So}, In_{So}\} \times \beta^{t+1}
\end{equation}

\begin{equation}\label{simSocf}
 simSoc = \alpha \times simSoc\{D_{So}\} + (1-\alpha) \times simSoc\{In_{So}\}
\end{equation}

\begin{algorithm}[!t]
  \caption{Proximity Similarity Calculation}
  \label{algoirhtm-sim}
  \begin{algorithmic}[1]
  \State $t=0$ // In current~ time~ slot
  \State $simPro = 0$
  \State $\beta = 0.8$
  \For {t=0 to i-1}
    \State $sum = 0$
    \For {common nodes between nodes A and B}
       \State //$degree$ is corresponding value in Ego
       \State $sum=sum+degree$
    \EndFor
    \State $simPro = simPro + sum \times \beta$
    \State $\beta = \beta \times \beta$
  \EndFor
  \State
  \Return $simPro$
  \end{algorithmic}
\end{algorithm}

\subsection{Forwarding Utility Function}

According to our previous descriptions, the similarities between nodes A and D with respect to the three social dimensions are calculated as $simPro_{A}$ , $simIns_{A}$, and $simSoc_{A}$. Similarly, $simPro_{B}$ , $simIns_{B}$, and $simSoc_{B}$ can be computed to indicate the similarities between nodes B and D. Accordingly, PIS combines the three dimensions to make the final forwarding decision.

First, the similarity deviation values from three social dimensions $simDevPro$, $simDevIns$, and $simDevSco$ among nodes A and B are calculated according to Equations \eqref{devPro}, \eqref{devIns}, and \eqref{devSco}, respectively. Then, the forwarding utility function $simPIS$ is defined using Equation \eqref{simPIS}, where $\rho + \sigma + \tau =1$.

\begin{equation}\label{devPro}
 simDevPro = \frac{simPro_A-simPro_B}{simPro_A+simPro_B}
\end{equation}
\begin{equation}\label{devIns}
 simDevIns = \frac{simIns_A-simIns_B}{simIns_A+simIns_B}
\end{equation}
\begin{equation}\label{devSco}
 simDevSco = \frac{simSco_A-simSco_B}{simSco_A+simSco_B}
\end{equation}

\begin{eqnarray}
\begin{split}\label{simPIS}
 simPIS &=& \rho \times simDevPro  \\
        & & + \sigma \times simDevIns \\
        & & + \tau \times simDevSoc
\end{split}
\end{eqnarray}

\subsection{The Importance of Social Parameters}

As introduced in the previous subsection, values $\rho$, $\sigma$, and $\tau$ are the tuning parameters in the PIS protocol that help adjust the significance of the three social dimensions which are highly dependent on the routing application. Since routing scenarios in real applications are generally complex and dynamic, identifying the importance of each social feature is a non-trivial problem. If the dominant social factors are known in a data routing context, users can assign reasonable importance to each tuning parameter to improve its performance. For instance, in a campus environment, students usually have regular mobility pattern and classmates have similar class schedules during weekdays. In this situation, the social relationships among the users (i.e., classmates) will be a dominant social feature in the message delivery. Hence, the value of parameter $\tau$ should be increased. However, in weekends, students usually get together for various activities based on their interests. In this situation, the user interests (i.e., parameter $\sigma$) should get a higher importance. As a result, parameters $\rho$, $\sigma$ and $\tau$ can affect a social-based routing protocol in different ways.

\subsection{Copy Control Mechanism}

The PIS protocol makes use of a copy control mechanism to control data congestion in the network. This idea is borrowed from the Spray-and-Wait protocol \cite{SprayandWait}. When a message is generated, an initialization value $nofCopy$ in assigned to the message to indicate the number of copies. When the message is forwarded, only half of the current value of $nofCopy$ is delivered. When all the message copies are disseminated and the value of $nofCopy$ for a message reaches to 1, the message can not be forwarded anymore. As a result, some forwarding opportunities might be missed. To tackle this problem, PIS introduces a range control parameter denoted as $\gamma$. In PIS, half of the value $nofCopy$ for a message can be forwarded to other nodes if $simPIS+\gamma~ \textgreater~ 0~ $, where $\gamma$ is assigned to expand the forwarding range. When a node has only one message copy, it forwards the message to another node if $simPIS~ \textgreater~ 0$. Then, it deletes the message from its buffer. Thus, the number of message copies is controlled using parameter $nofCopy$.

In summary, the forwarding strategy is outlined as follows:

1) If node A has more than one message copies and $simPIS+\gamma~ \textgreater~ 0~ $, then node A forwards half number of current copies to node B. $\gamma$ is assigned to prevent messages forwarded too early to limit the forwarding range.

2) When node A has only one message copy and $simPIS \textgreater 0$, it forwards message to node B and deletes the message copy from itself.

The pseudo-code of PIS algorithm is presented in Algorithm \ref{algorithm-routing}.

\begin{algorithm}[!t]
  \caption{PIS Algorithm}
  \label{algorithm-routing}
    \begin{algorithmic}[1]
    \State Nodes A and B contact each other
    \State Upon reception of Hello message from node B do
    \State Exchange information of contact list, $S_{I}$ and $D_{So}$
    \State Update $Ego$, $C_{I}$ and $In_{So}$
    \State Exchange message list
    \For {$M$ in the list of messages in node B}
      \State Compute $simPro_{A}$ , $simIns_{A}$ and $simSoc_{A}$
      \State Compute $simDevPro$, $simDevIns$ and
      \State ~~~~~~~~~~~~~$simDevSco$
      \State Compute $simPIS$
      \If {$nofCopy~of~M$ \textgreater 1}
            \If {$simPIS+\gamma \textgreater 0$}
                \State $nofCopy$=$cofCopy$/2
                \State transferMsgs.add($M$, connection)
            \EndIf
      \ElsIf {$nofCopy~of~M$==1}
            \If {$simPIS$ \textgreater 0}
                \State transferMsgs.add($M$, connection)
                \State deleteMsg from node B when transfer done
            \EndIf
      \EndIf
    \EndFor
  \end{algorithmic}
\end{algorithm}

\section{Performance Evaluation}

We compare the performance of the PIS protocol using Opportunistic Network Environment (ONE) \cite{a-one} simulator which is a trace driven simulator particularly for intermittently connected networks.

\subsection{Datasets}

In this section, we evaluate the performance of the PIS protocol using two real traces: SIGCOMM09 \cite{sigcomm09} and INFOCOM06 \cite{scott2009crawdad}. The SIGCOMM09 data set was collected using an opportunistic mobile social application MobiClique during SIGCOMM 2009 conference. In this data set, around 100 smart phones were distributed among a set of volunteers during two days. During this period, the traces of Bluetooth proximities, opportunistic message creation and dissemination, and activity period of the participants were recorded. In addition, each participant was asked to log on to their Facebook profile in order to capture the list of her Facebook friends and interests. The proximity information of the users was captured every 120 $\pm$ 10 seconds.

We also use INFOCOM06 data set which was collected during IEEE INFOCOM 2006 conference. In this data set, 78 participants were asked to carry iMotes device during the conference in order to collect their opportunistic contacts and mobility statistics. Additionally, some personal information of the participants such as their interests, language, and affiliation were also collected through a survey questionnaire which represented their individual and social attributes.

In above data sets, the physical proximity information can be obtained by analyzing the nodes' contact records. While, the interest information of the nodes are stored in the data sets directly. For the social relationship information, we use an explicit social relationship attribute in SIGCOMM09 data set which is called \textit{friendship}. In INFOCOM06 data set, on the other hand, there are several kinds of social relationships. For example, if the participants in a conference with the same language communicate with each other frequently, we use \textit{language relationship} in this data set to establish the social relationships among the nodes.

\subsection{Simulation Setup}

We compare the effectiveness of PIS with three well-known routing protocols, such as Epidemic routing \cite{Epidemic}, PROPHET \cite{PROPHET}, and SimBet \cite{Routing:Daly}. In the PROPHET, the next intermediate nodes are selected using their contact history. The Epidemic routing adopts a simple flooding method in which each node copies messages in her buffer to other encountered nodes if they have not received them yet. The SimBet utilizes the betweenness centrality and similarities of nodes to choose the next message carriers based on the ego network concept. In the SimBet algorithm, the value of the ego matrix is set to 0 if there is no contact and 1 if there exist a contact. While, in PIS, the value of ego network matrix indicates the contact frequency of the nodes.

In the simulations, four performance metrics are evaluated as followings:

\begin{itemize}
\item\textbf{Delivery Ratio}: the ratio of successfully delivered messages to the total number of unique messages created within a given period.
\item\textbf{Overhead Ratio}: the ratio of relayed messages and delivered messages, reflecting the ratio of message replicas propagated into the network.
\item\textbf{Average Latency}: the average time between the time a message is generated and the time it is delivered successfully, including buffering delays.
\item\textbf{Average Hop Count}: the average hop-counts when messages are received successfully.

\end{itemize}

We run each simulation setting 30 times and calculate the average values. The simulation parameters are summarized in Table \ref{simPrameter}.

\begin{table}[!b]
\centering
\caption{Simulation parameters}
\label{simPrameter}
\begin{tabular}{c c }                                                  \hline
Simulation Parameters       & Values                 \\ \hline
Duration period             & 40 hours                 \\ 
Warm up                     & 5000 (second)         \\ 
Nodes' speed                & 0.5$\scriptsize{\sim} $1.5~ (m/s)       \\ 
Wait time at destination    & 100$\scriptsize{\sim} $200~ (second)       \\ 
Interface Type              &Bluetooth               \\ 
Transmit speed              &250 KB               \\ 
Transmit range              &10 meter                 \\ 
Moment model                &External Movement             \\ 
Data set                    &SIGCOMM09, INFOCOM06          \\ 
Event interval              & 500$\scriptsize{\sim}$650         \\
Message size                & 500$\scriptsize{\sim}$1024 (MB)         \\ 
Message TTL                 & 10 (hour)      \\
Nodes' buffer               & 5 (MB)         \\
time slot parameter $i$     & 6                         \\
$\alpha$                    & 0.5                   \\
$\beta$                     & 0.8                   \\ 
$\gamma$                    & 0.8, 0.1                    \\
$\rho$, $\sigma$ and $\tau$ & 1/3, 1/3, and 1/3             \\    \hline
\end{tabular}
\end{table}

\begin{figure*}[t]
\centering
\subfigure[Delivery Ratio]{
\label{sig_perf_delivery}
\includegraphics[width=0.23\textwidth]{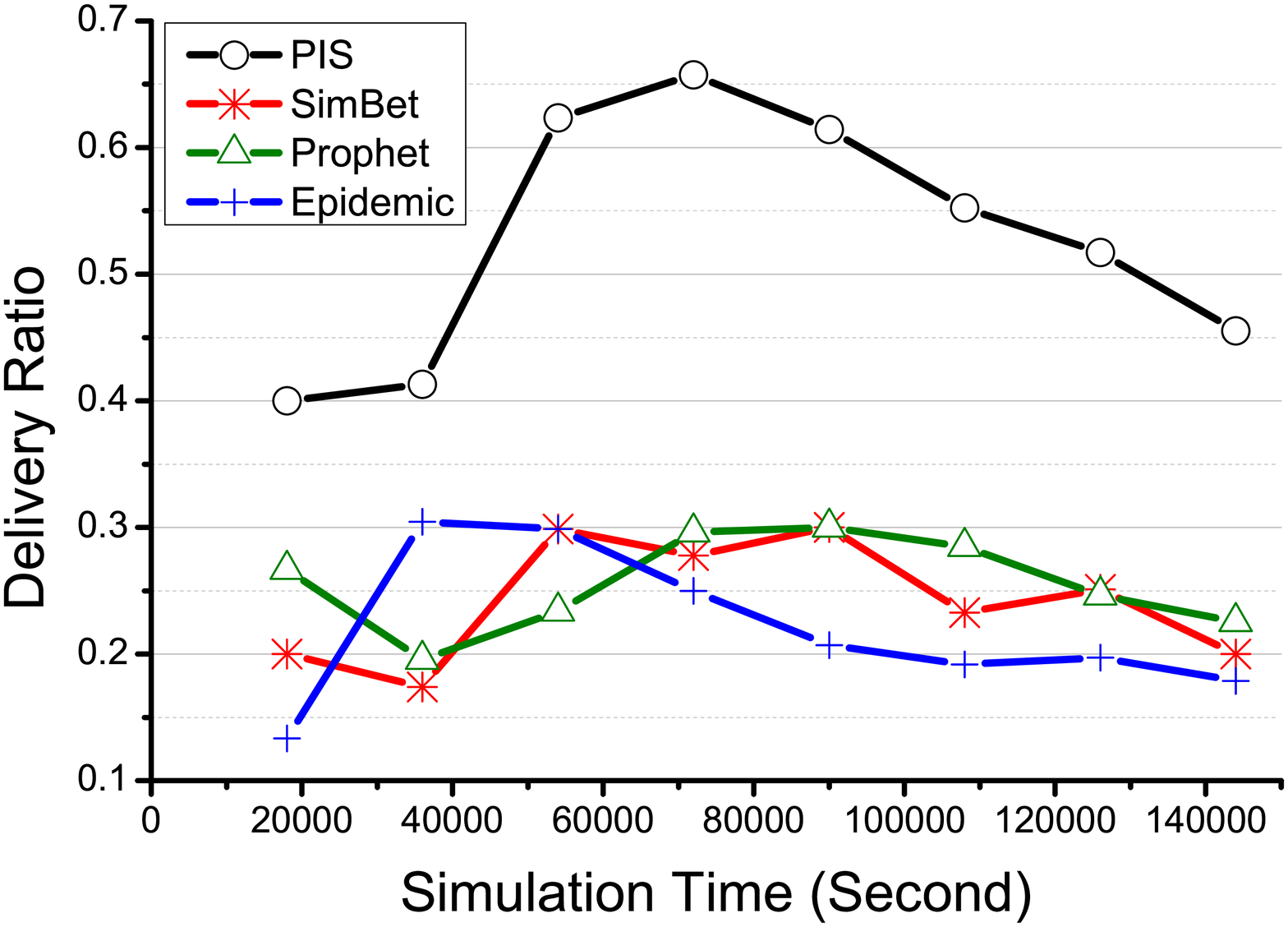}}
\subfigure[Overhead]{
\label{sig_perf_overhead}
\includegraphics[width=0.23\textwidth]{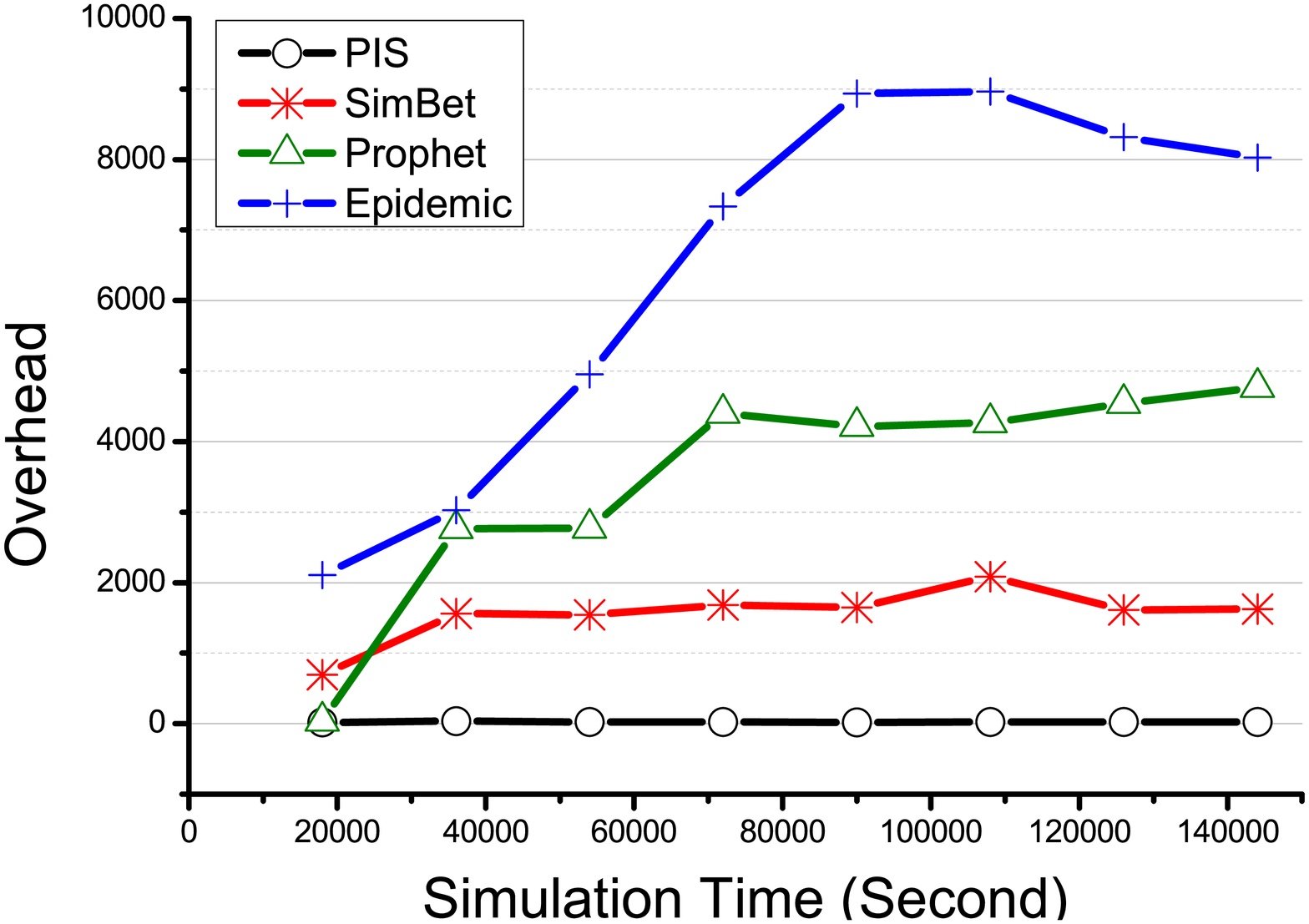}}
\subfigure[Average Latency]{
\label{sig_perf_latency}
\includegraphics[width=0.23\textwidth]{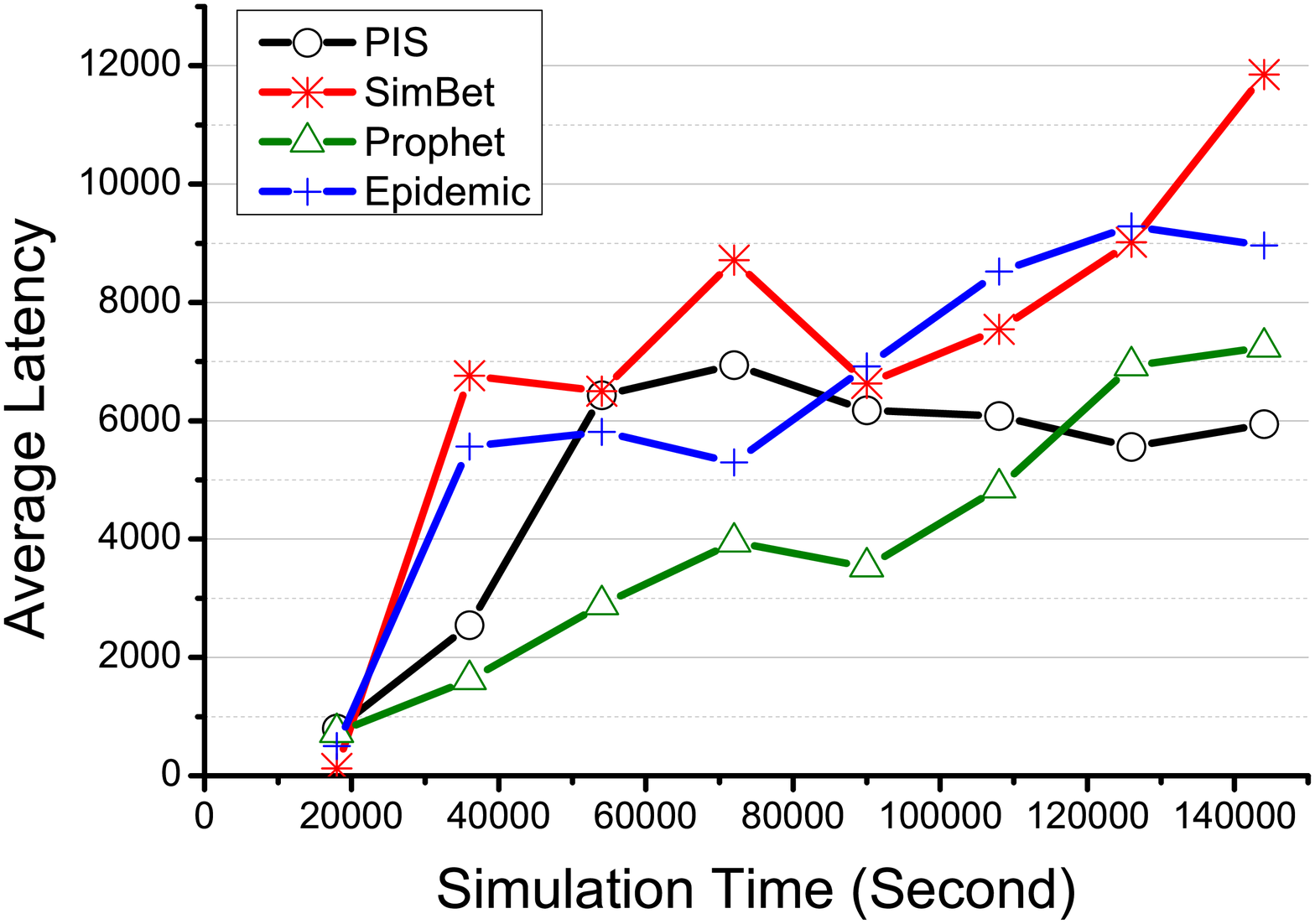}}
\subfigure[Average Hop Count]{
\label{sig_perf_hop}
\includegraphics[width=0.23\textwidth]{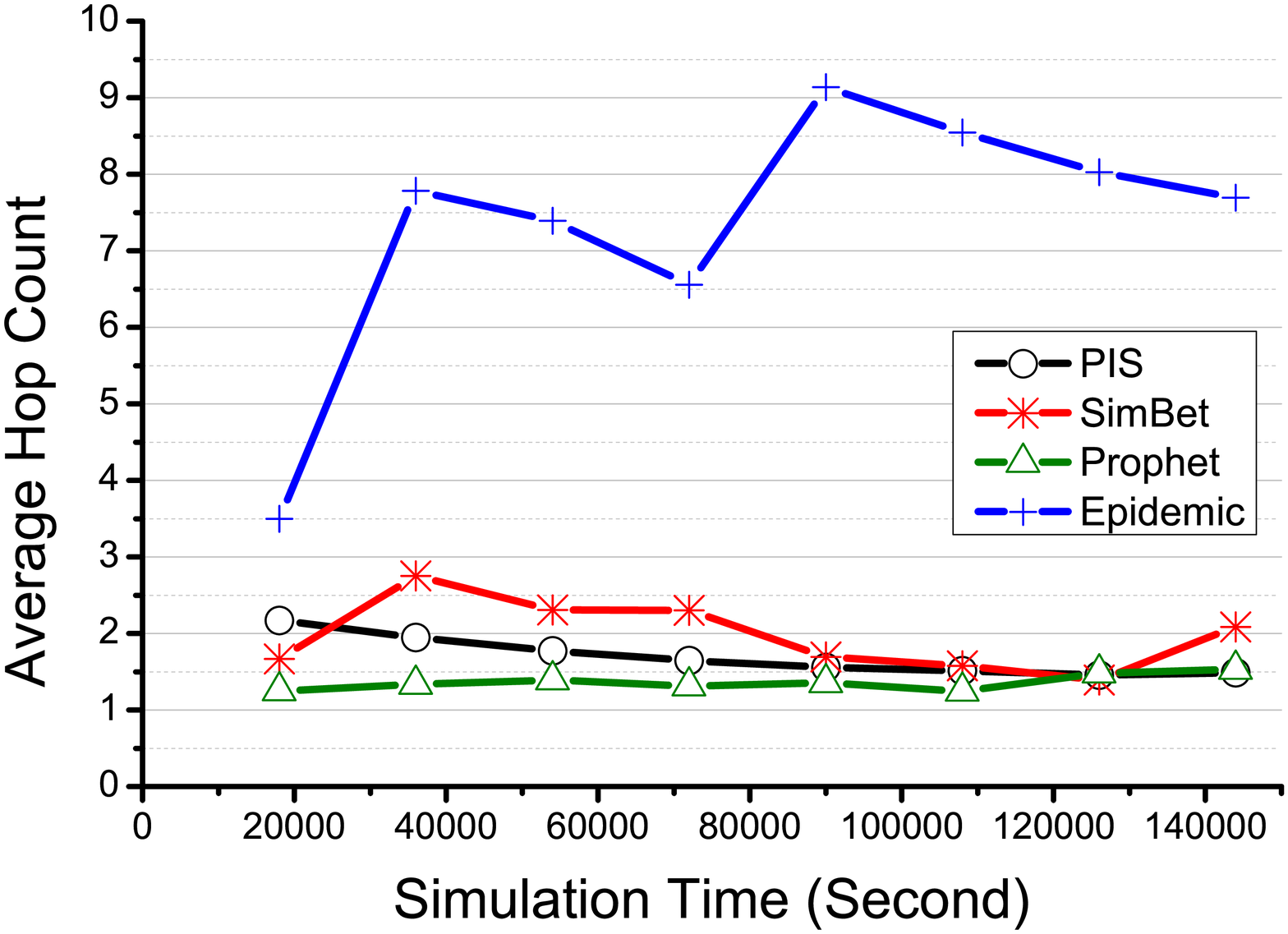}}
\caption{The comparison of PIS with the other routing algorithms using SIGCOMM09 data set.}
\label{per-sig}
\end{figure*}

\begin{figure*}[t]
\centering
\subfigure[Delivery Ratio]{
\label{info_perf_delivery}
\includegraphics[width=0.23\textwidth]{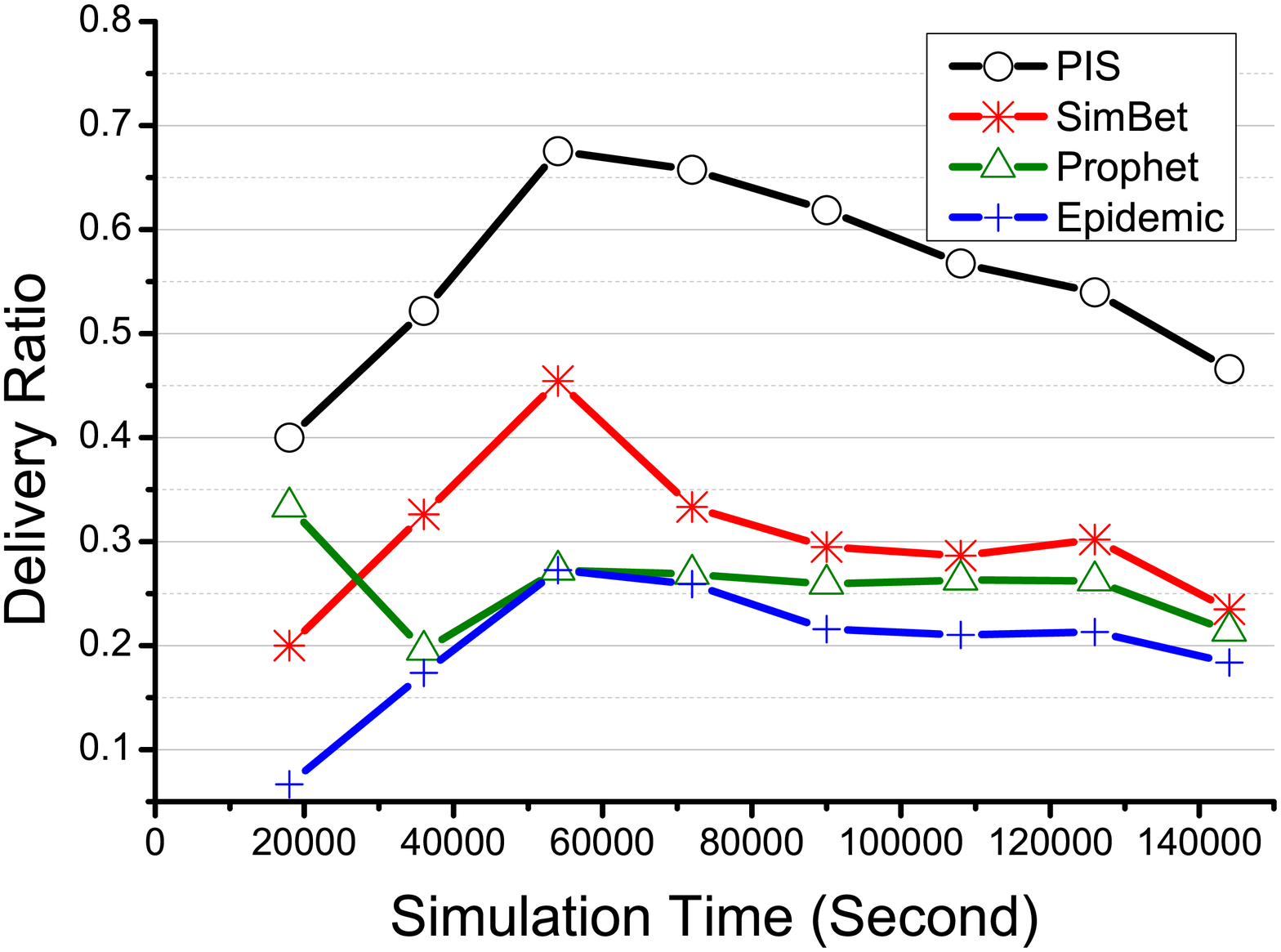}}
\subfigure[Overhead]{
\label{info_perf_overhead}
\includegraphics[width=0.23\textwidth]{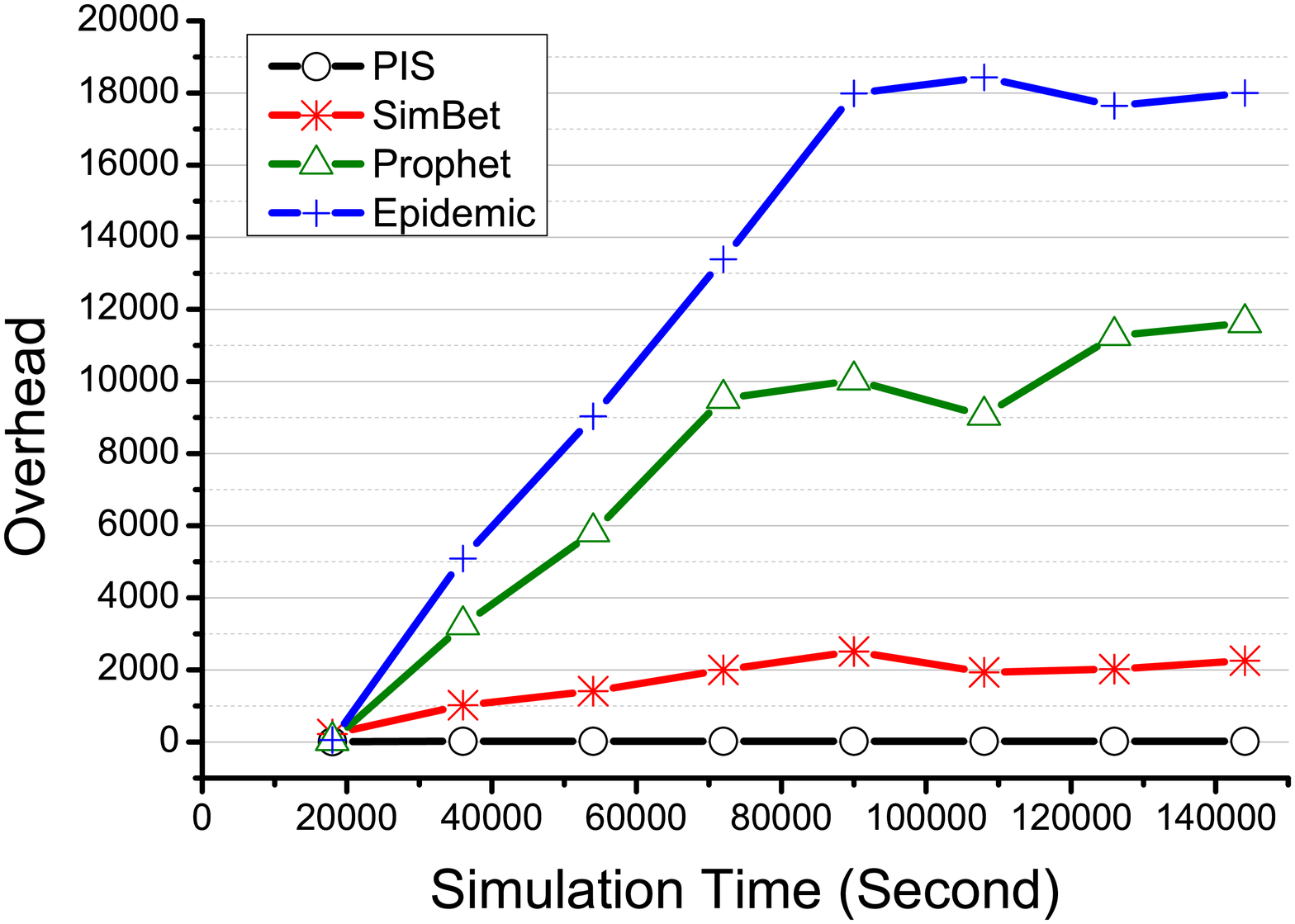}}
\subfigure[Average Latency]{
\label{info_perf_latency}
\includegraphics[width=0.23\textwidth]{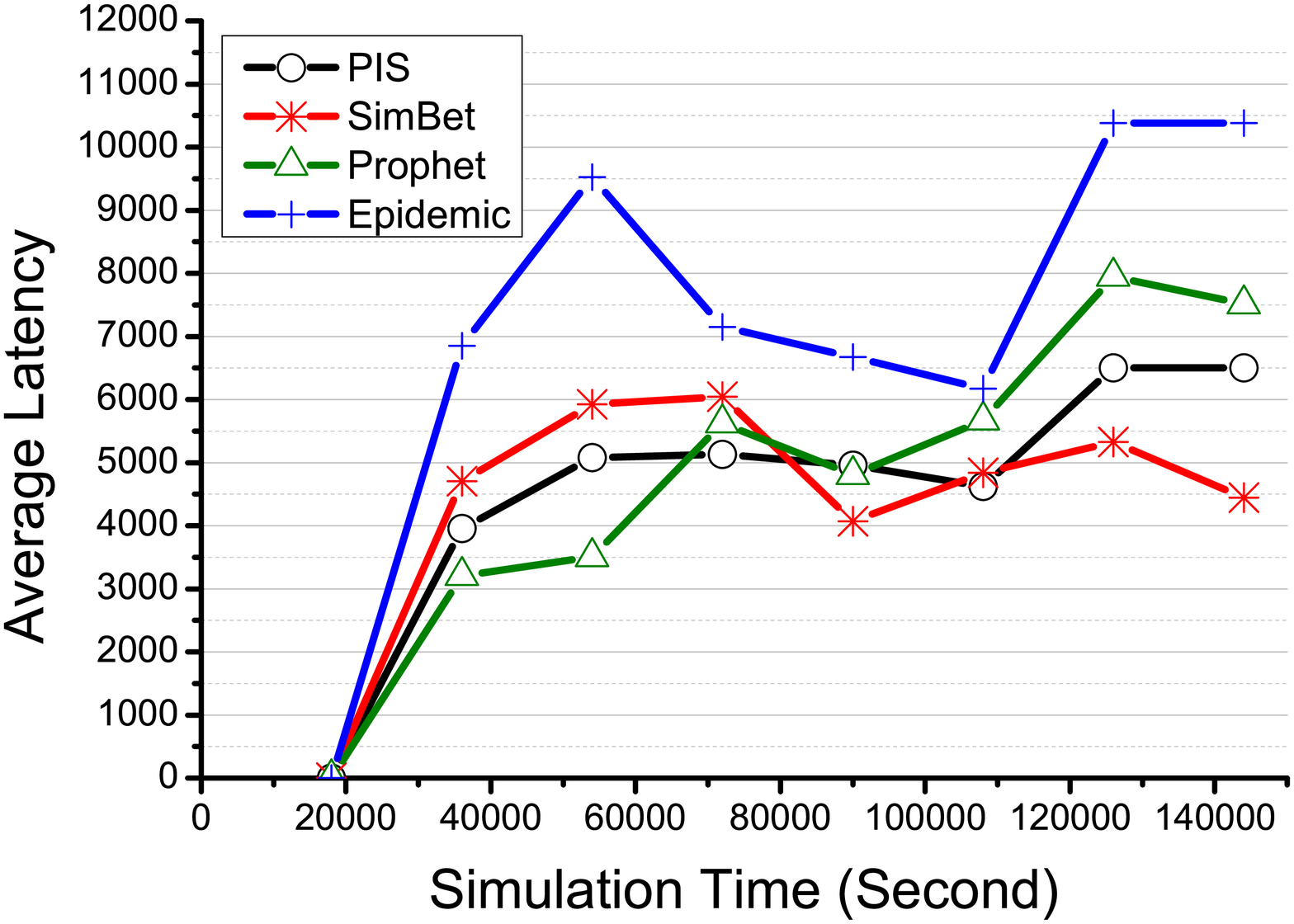}}
\subfigure[Average Hop Count]{
\label{info_perf_hop}
\includegraphics[width=0.23\textwidth]{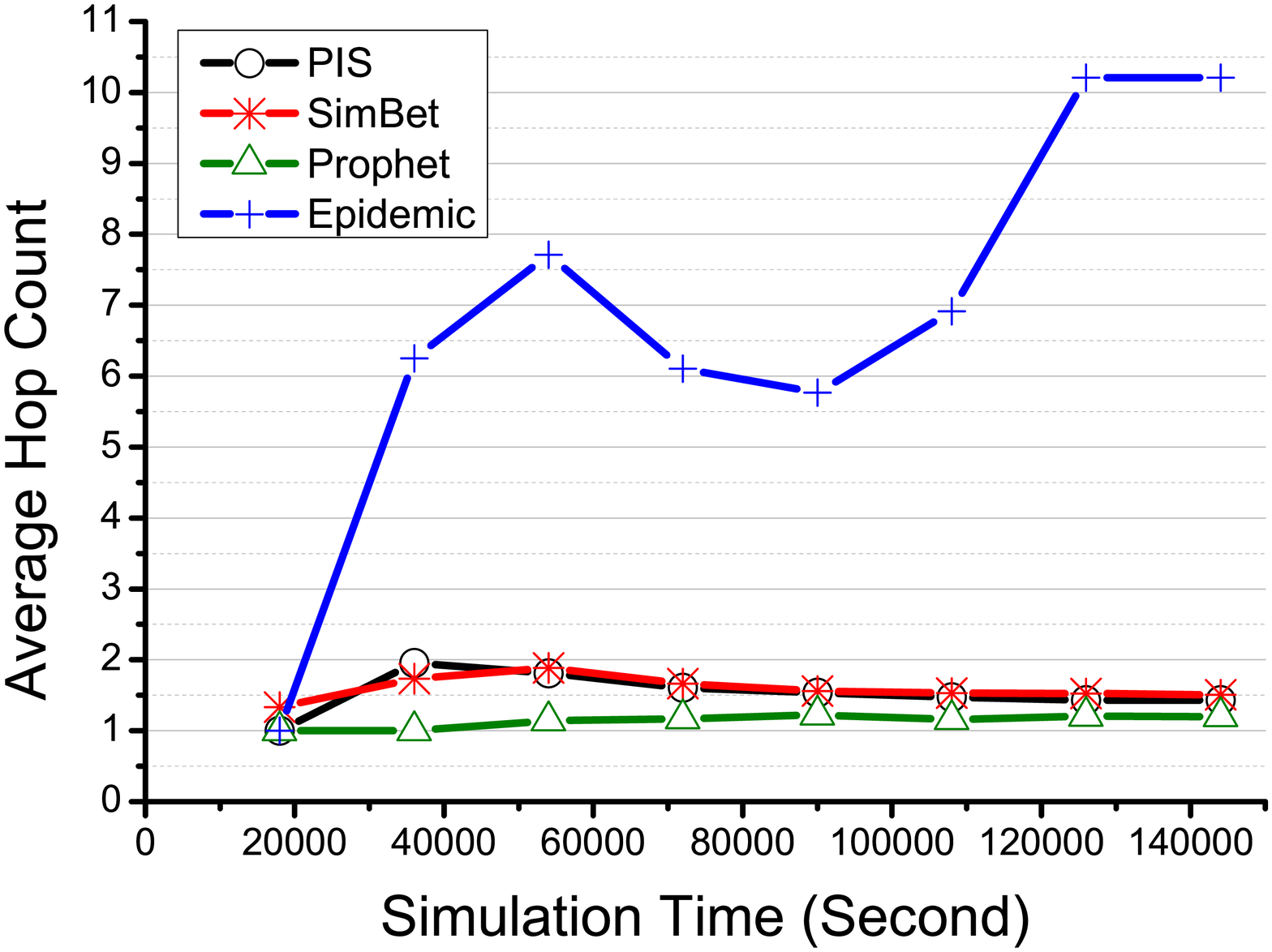}}
\caption{Efficiency comparison of PIS with the other routing algorithms using INFOCOM06 data set.}
\label{per-info}
\end{figure*}

\subsection{Experimental Results and Analysis}

\subsubsection{Efficiency Comparison}

In Figs. \ref{per-sig} and \ref{per-info}, we show the performance of the algorithms in terms of delivery ratio, overhead ratio, average latency, and average hop count using SIGCOMM09 and INFOCOM06 data sets, respectively.

\begin{figure*}[!t]
\centering
\subfigure[Delivery Ratio]{
\label{gamma-delivery}
\includegraphics[width=0.23\textwidth]{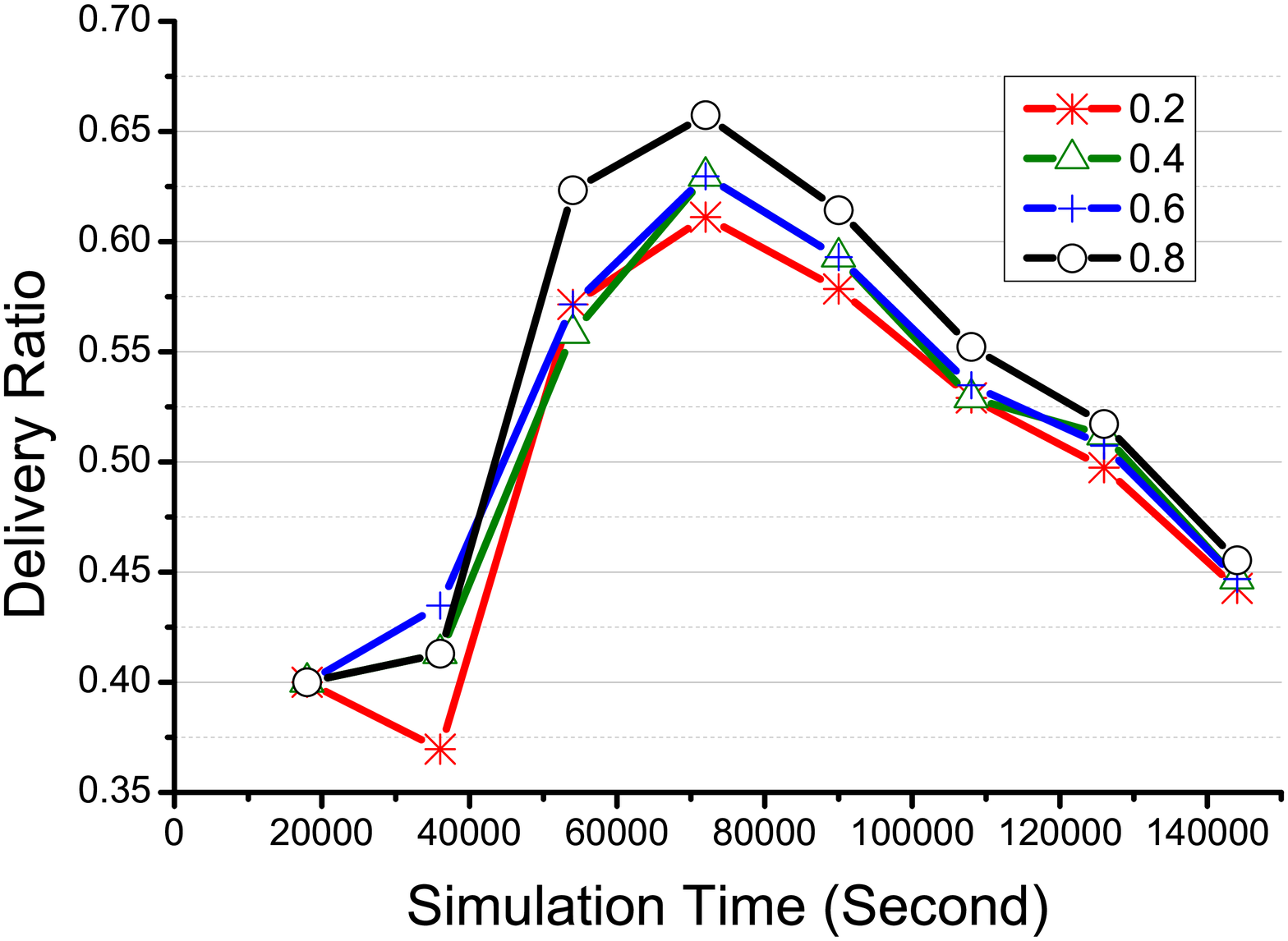}}
\subfigure[Overhead]{
\label{gamma-overhead}
\includegraphics[width=0.23\textwidth]{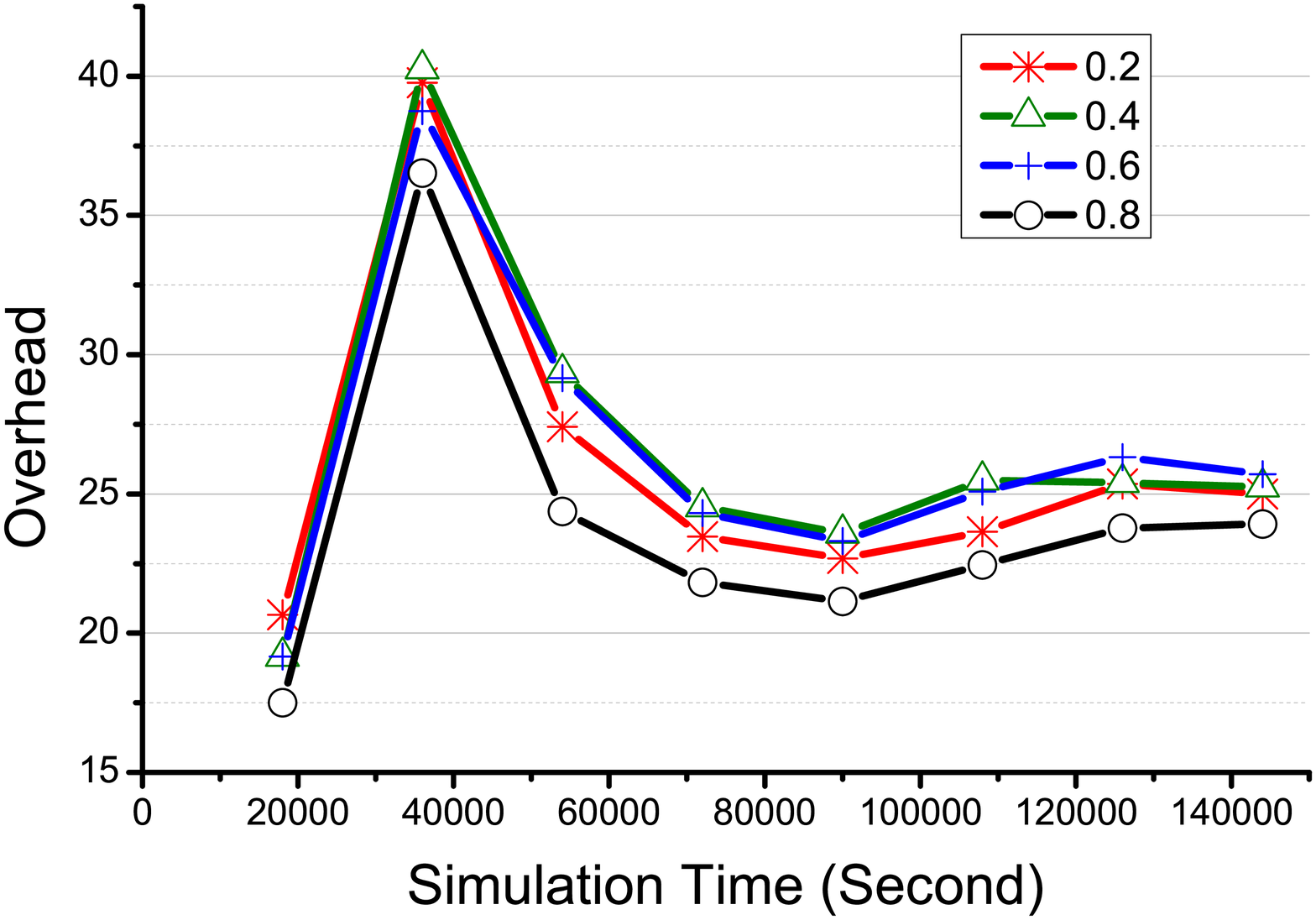}}
\subfigure[Average Latency]{
\label{gamma-latency}
\includegraphics[width=0.23\textwidth]{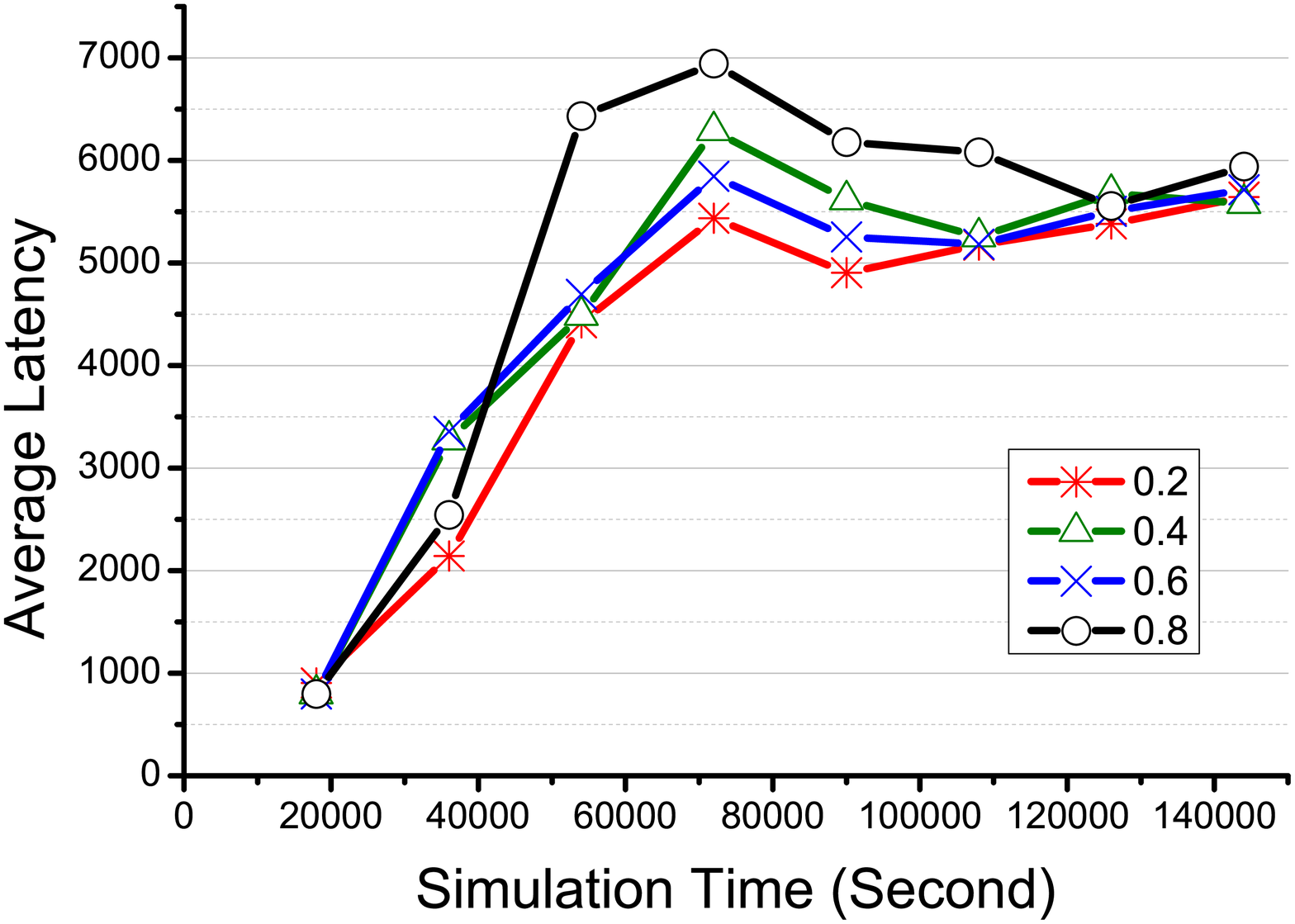}}
\subfigure[Average Hop Count]{
\label{gamma-hop}
\includegraphics[width=0.23\textwidth]{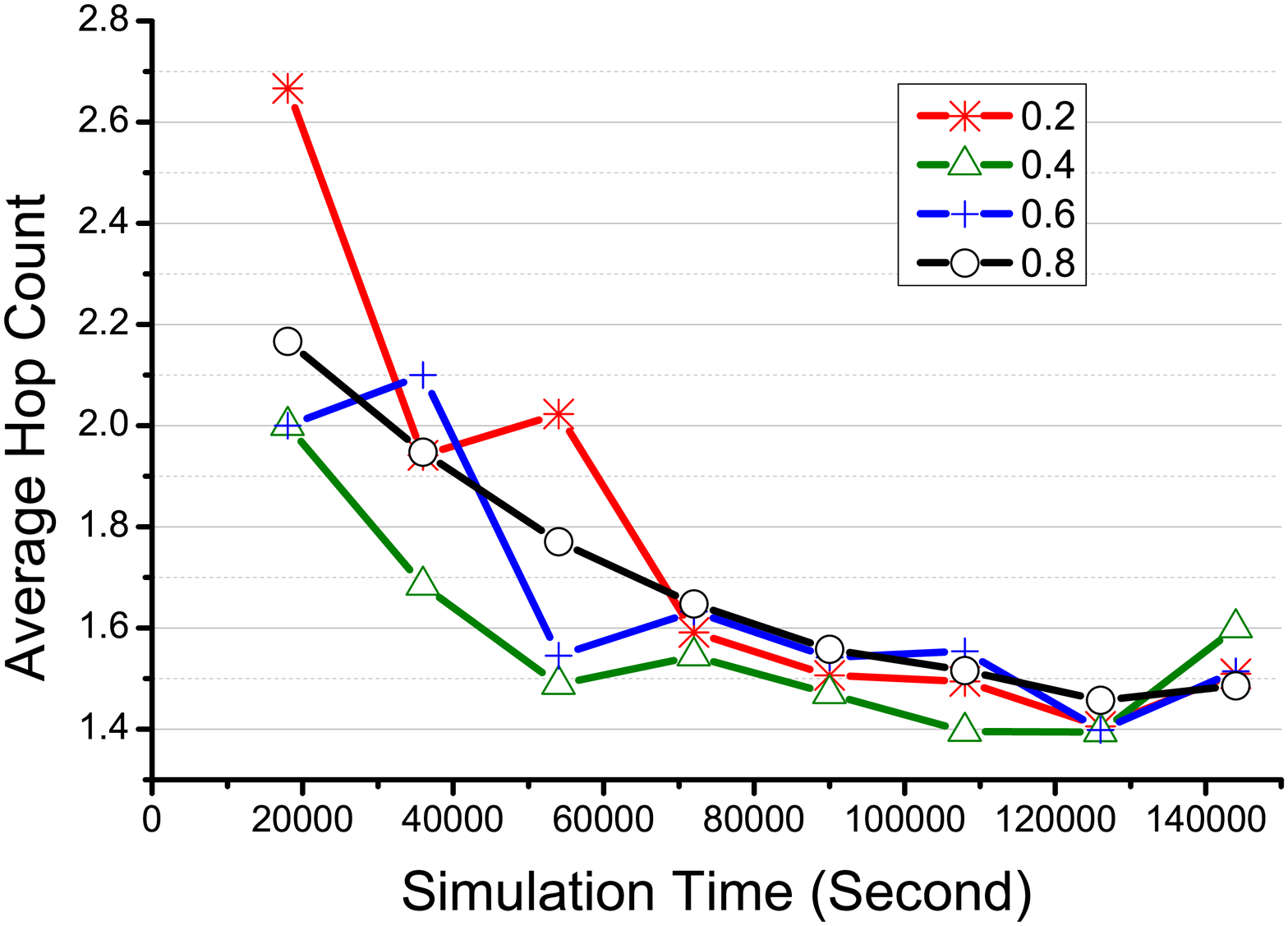}}
\caption{Simulation results for PIS under different $\gamma$ using SIGCOMM09 data set.}
\label{gamma-sig}
\end{figure*}

As it can be seen, PIS outperforms the other three protocols which have the highest delivery ratio and the lowest overhead. For example at time 20 hours in Fig. \ref{per-sig}, PIS forwards 65.74\% messages, while the delivery ratio of SimBet, PROPHET, and Epidemic are 27.78\%, 29.63\%, and 25\%, respectively. The overhead ratio of PIS is 21.83 which is far lower than SimBet with 1684, PROPHET with 4403, and Epidemic with 7336. Furthermore, the average latency of PIS is 6943 seconds which is smaller than SimBet with 8717 seconds, and higher than PROPHET with 3948 seconds, and Epidemic with 5296 seconds. The average hop count of PIS is 1.64 as compared to 2.3 for SimBet , 1.31 for PROPHET, and 6.5 for Epidemic .

\begin{figure*}[!t]
\centering
\subfigure[Delivery Ratio]{
\label{gamma-info-delivery}
\includegraphics[width=0.23\textwidth]{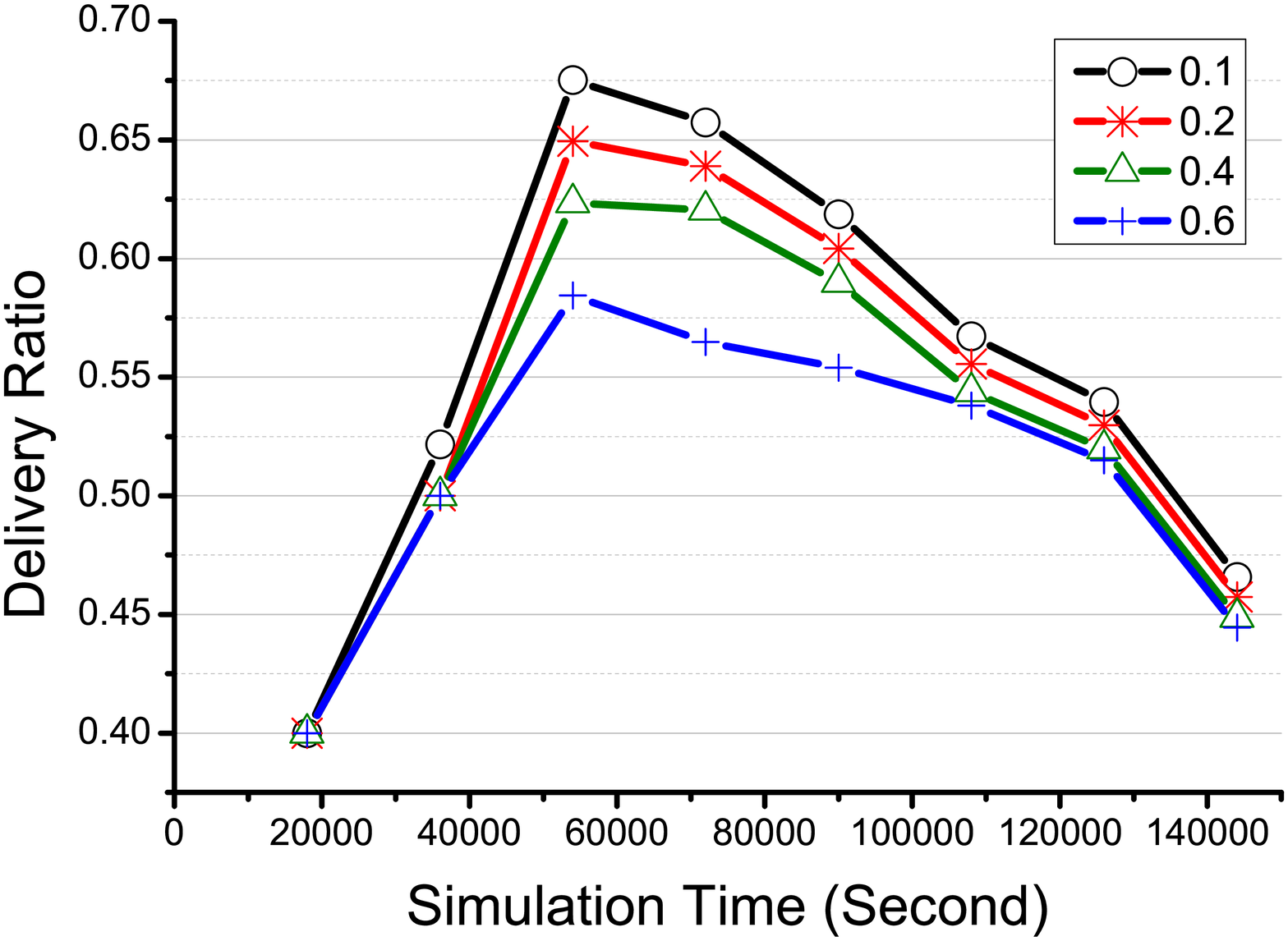}}
\subfigure[Overhead]{
\label{gamma-info-overhead}
\includegraphics[width=0.23\textwidth]{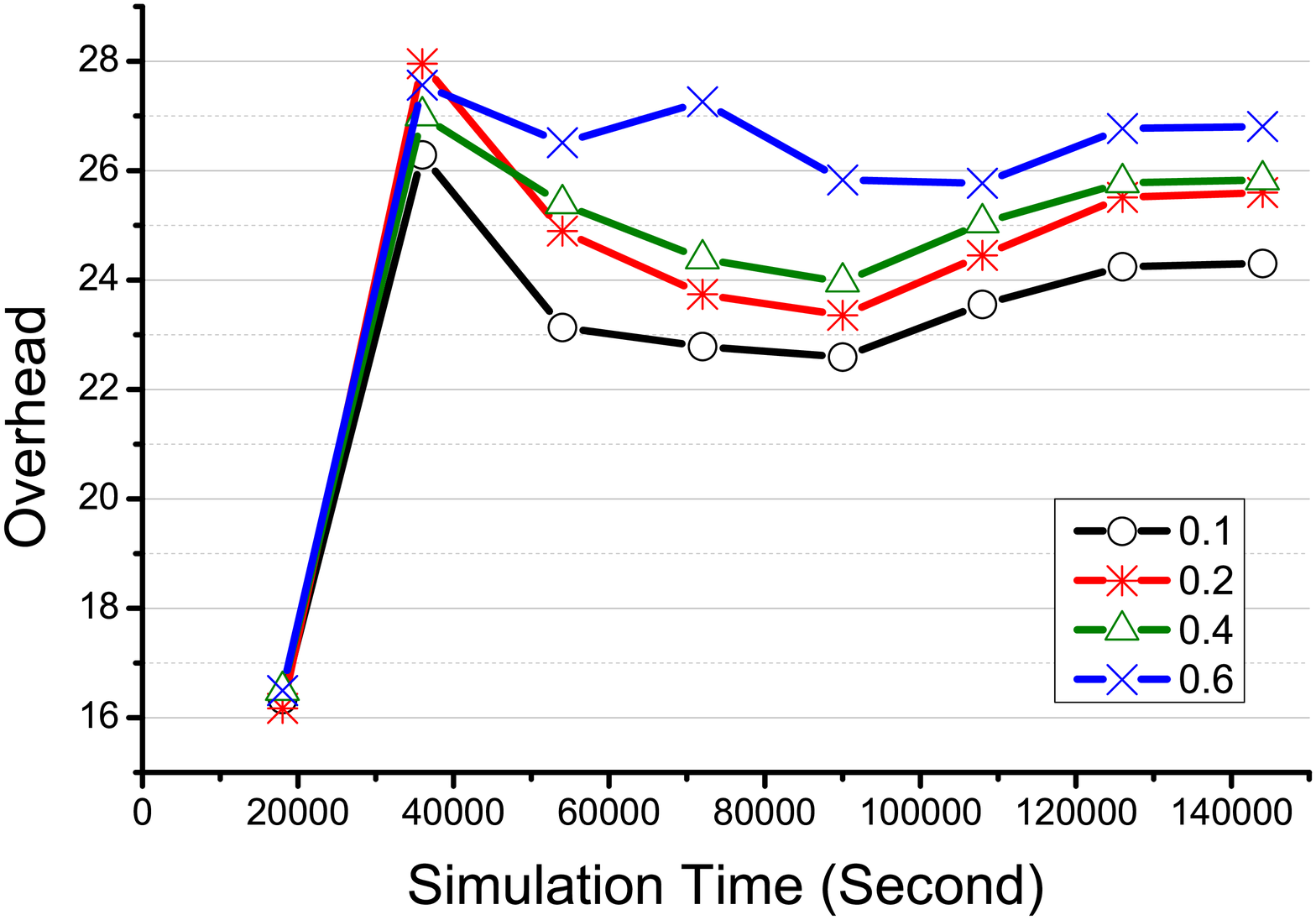}}
\subfigure[Average Latency]{
\label{gamma-info-latency}
\includegraphics[width=0.23\textwidth]{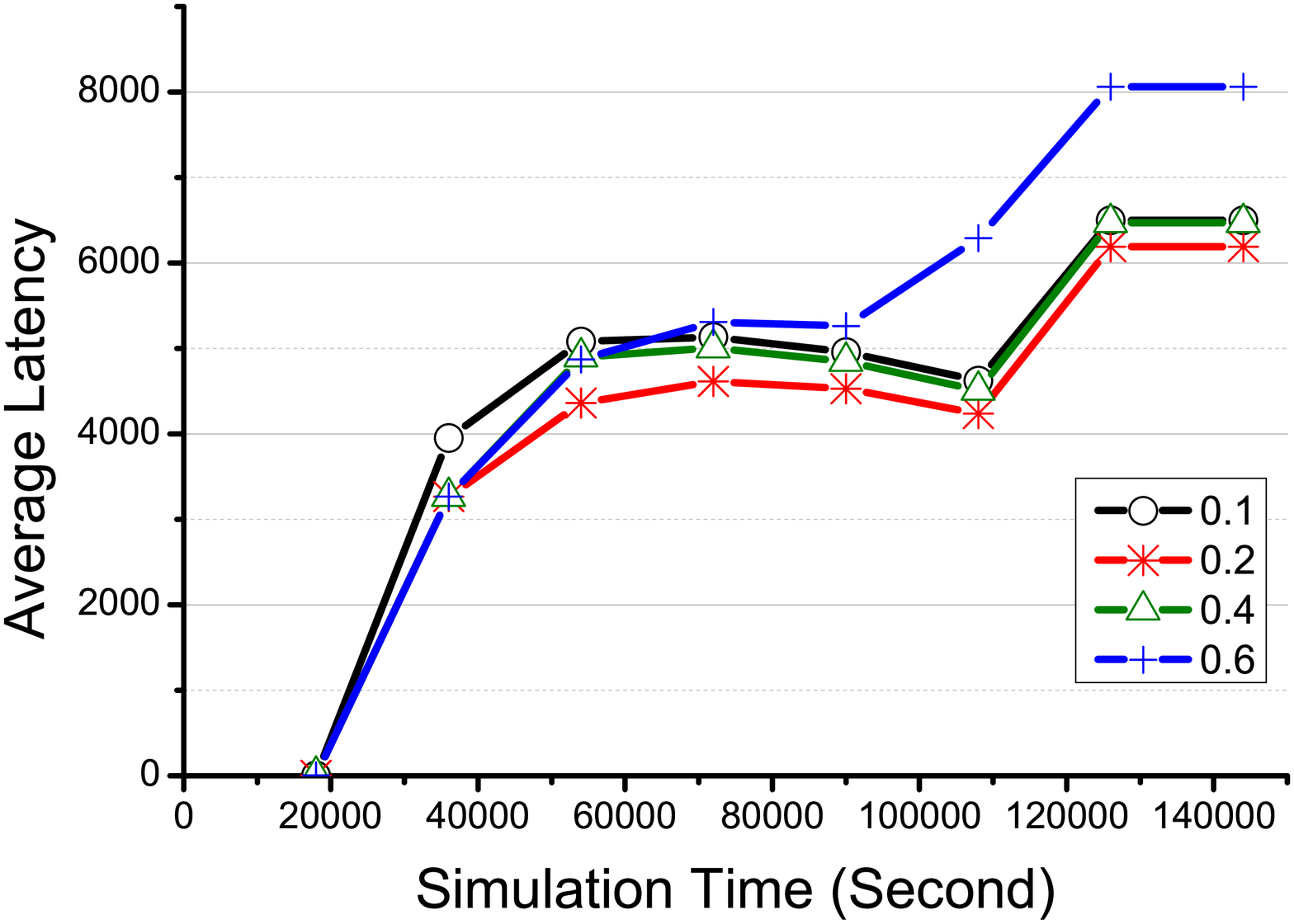}}
\subfigure[Average Hop Count]{
\label{gamma-info-hop}
\includegraphics[width=0.23\textwidth]{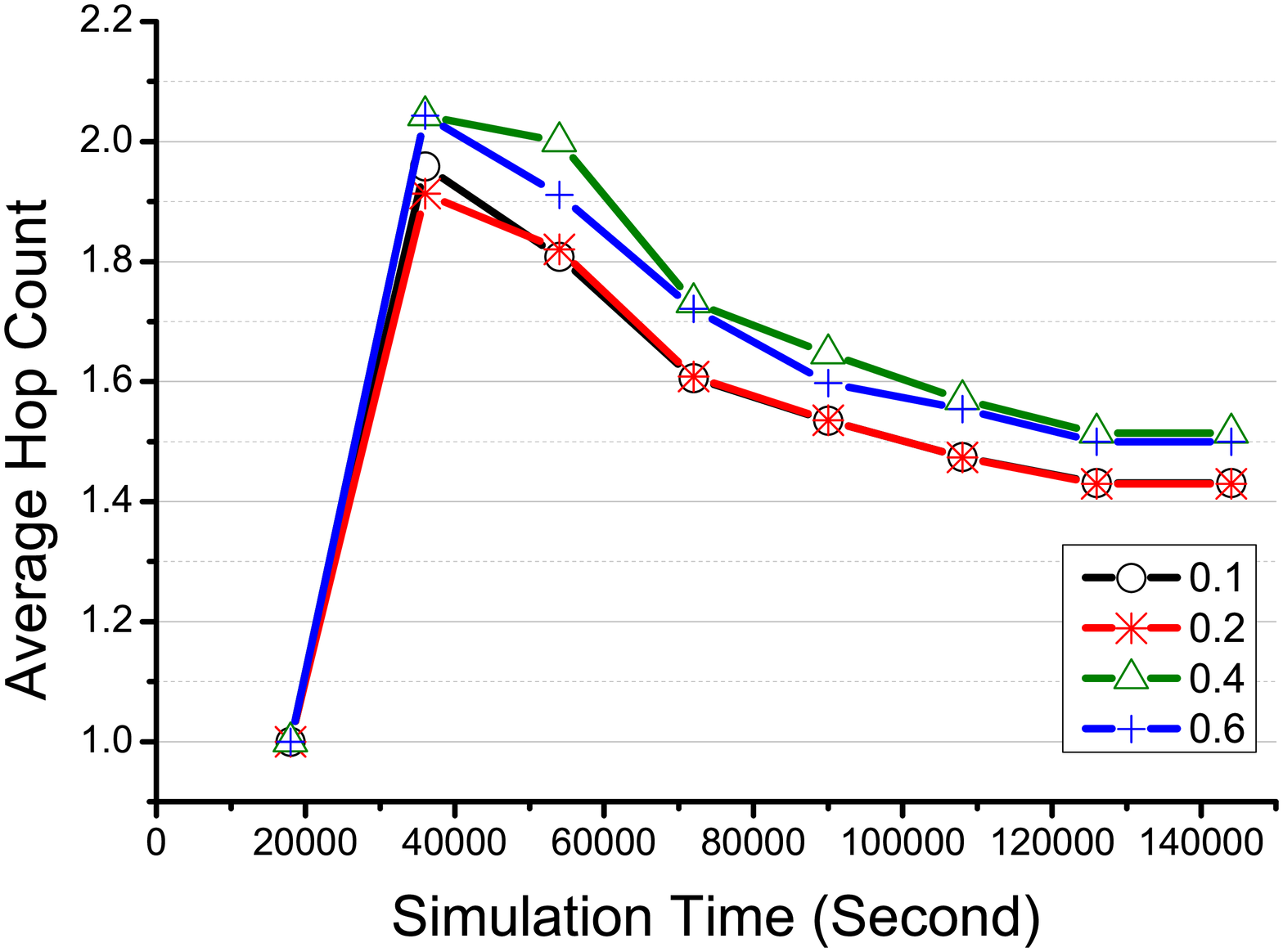}}
\caption{Simulation results for PIS under different $\gamma$ using INFOCOM06 data set.}
\label{gamma-info}
\end{figure*}

In INFOCOM06 data set, PIS achieves the best efficiency, as shown in Fig. \ref{per-info}. At 15 hours, the delivery ratio of PIS reaches 67.53\% which is 22\%, 40\%, and, 40\%  higher than SimBet, PROPHET, and Epidemic, respectively. The overhead ratio of PIS is 23 which is far lower than SimBet with 1409, PROPHET with 5826, and Epidemic with 9030. The average latency of PIS is 5079s which are smaller than SimBet with 5923 seconds and Epidemic with 9525 seconds, but longer than PROPHET with 3507 seconds. However, the average hop count of PIS is 1.8 where those for SimBet, PROPHET, and Epidemic are 1.8, 1.1, and 7.7, respectively.

The main reason that PIS outperforms the other protocols is that it integrates the three social features with time regularity to predict the future contacts of the nodes which are more reasonable and stable to select the appropriate intermediate nodes as forwarders. Additionally, PIS adopts the copy control mechanism which decreases the routing overhead considerably. In PIS protocol, when the utility function $simPIS$ is higher than 0 and the number of message copies is 1, the message is forwarded from a low-value node to a high-value node. At the same time, the value of message copy in the node with the low-value node is reset when the message is forwarded successfully. This method not only controls the number of message copies but also improves the forwarding opportunities.

\begin{figure*}[!t]
\centering
\subfigure[Delivery Ratio]{
\label{interval-delivery}
\includegraphics[width=0.23\textwidth]{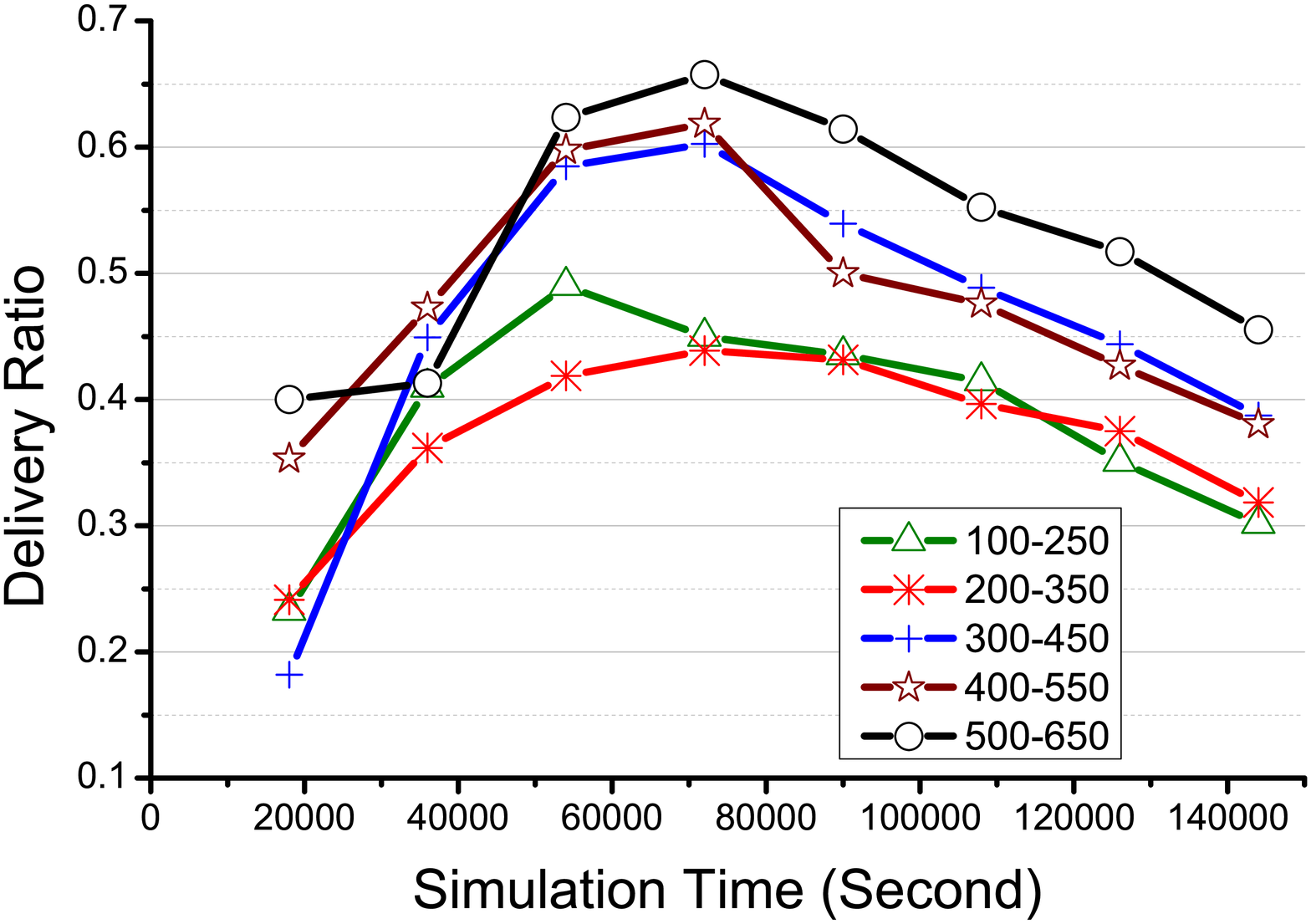}}
\subfigure[Overhead]{
\label{interval-overhead}
\includegraphics[width=0.23\textwidth]{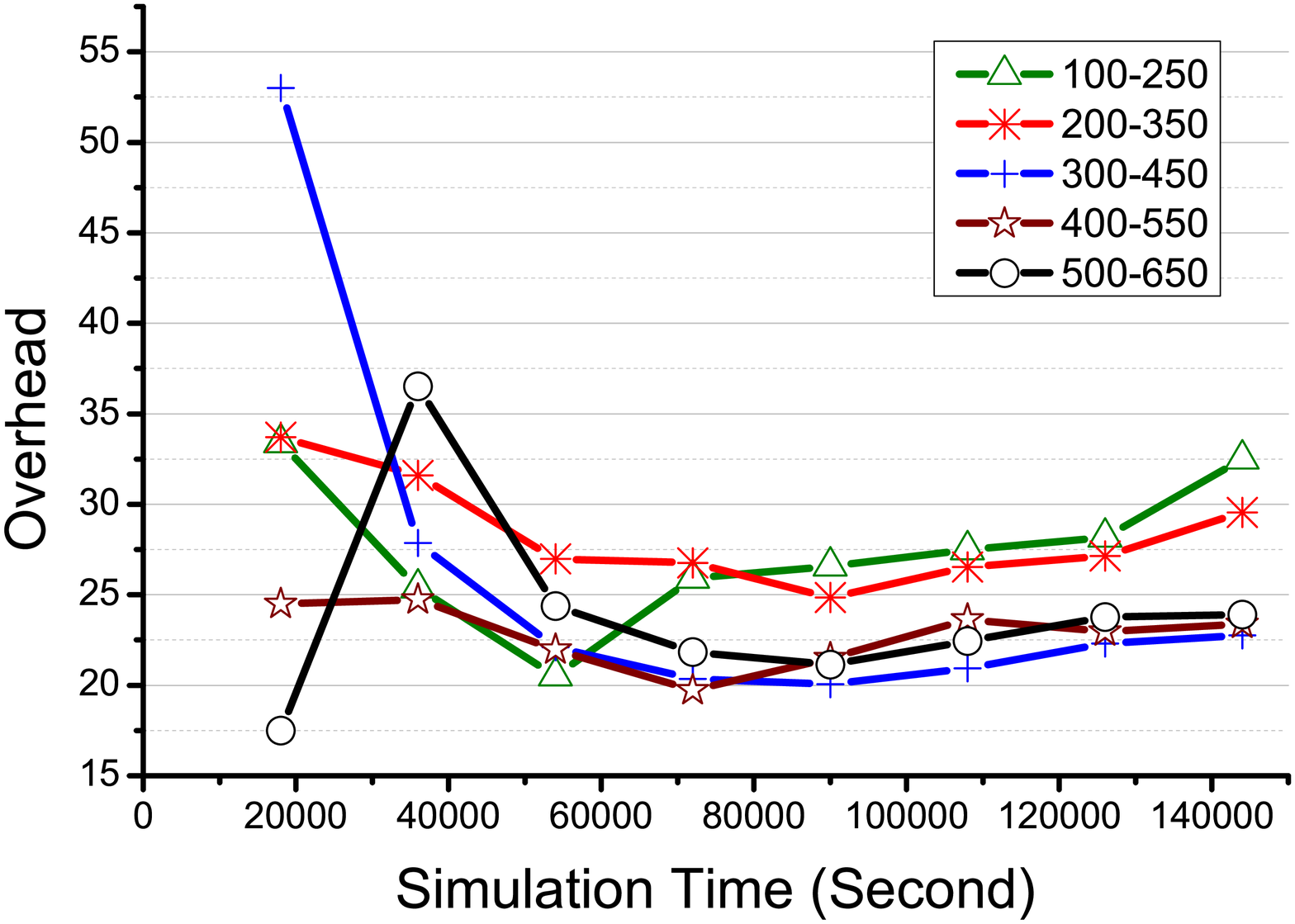}}
\subfigure[Average Latency]{
\label{interval-latency}
\includegraphics[width=0.23\textwidth]{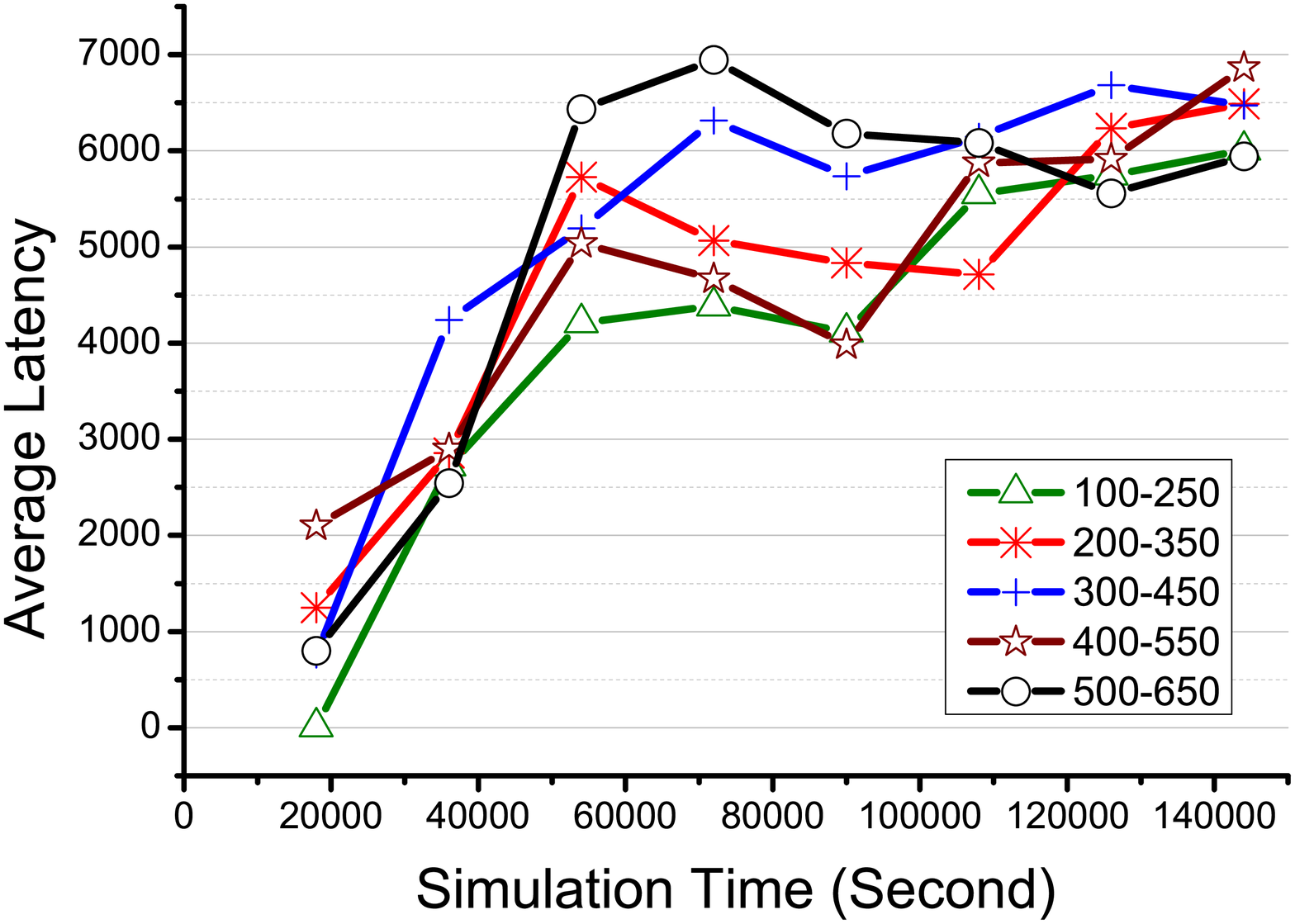}}
\subfigure[Average Hop Count]{
\label{interval-hop}
\includegraphics[width=0.23\textwidth]{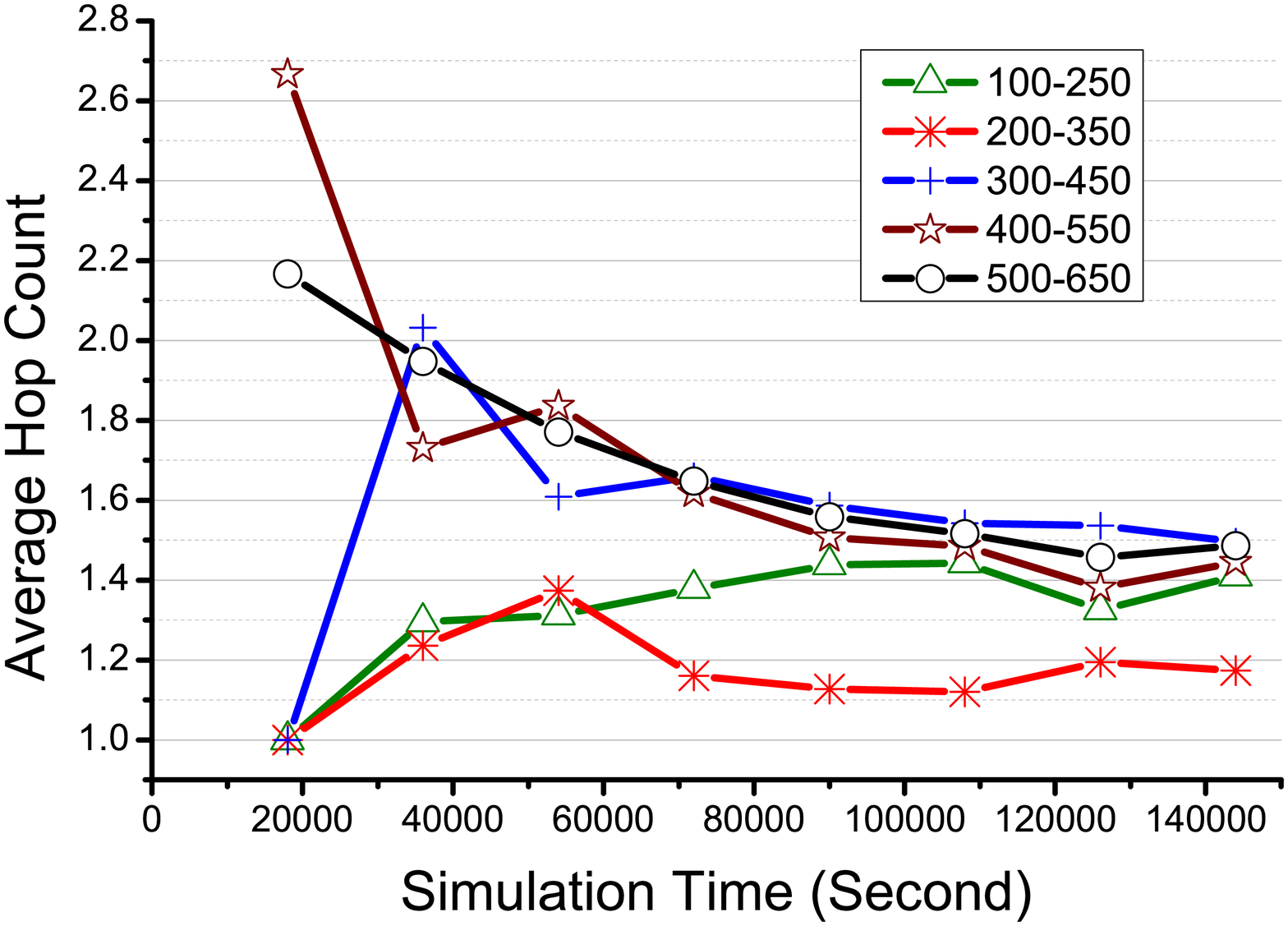}}
\caption{Simulation results for PIS under different message intervals using SIGCOMM09 data set.}
\label{interval-sig}
\end{figure*}

\begin{figure*}[!t]
\centering
\subfigure[Delivery Ratio]{
\label{interval-info-delivery}
\includegraphics[width=0.23\textwidth]{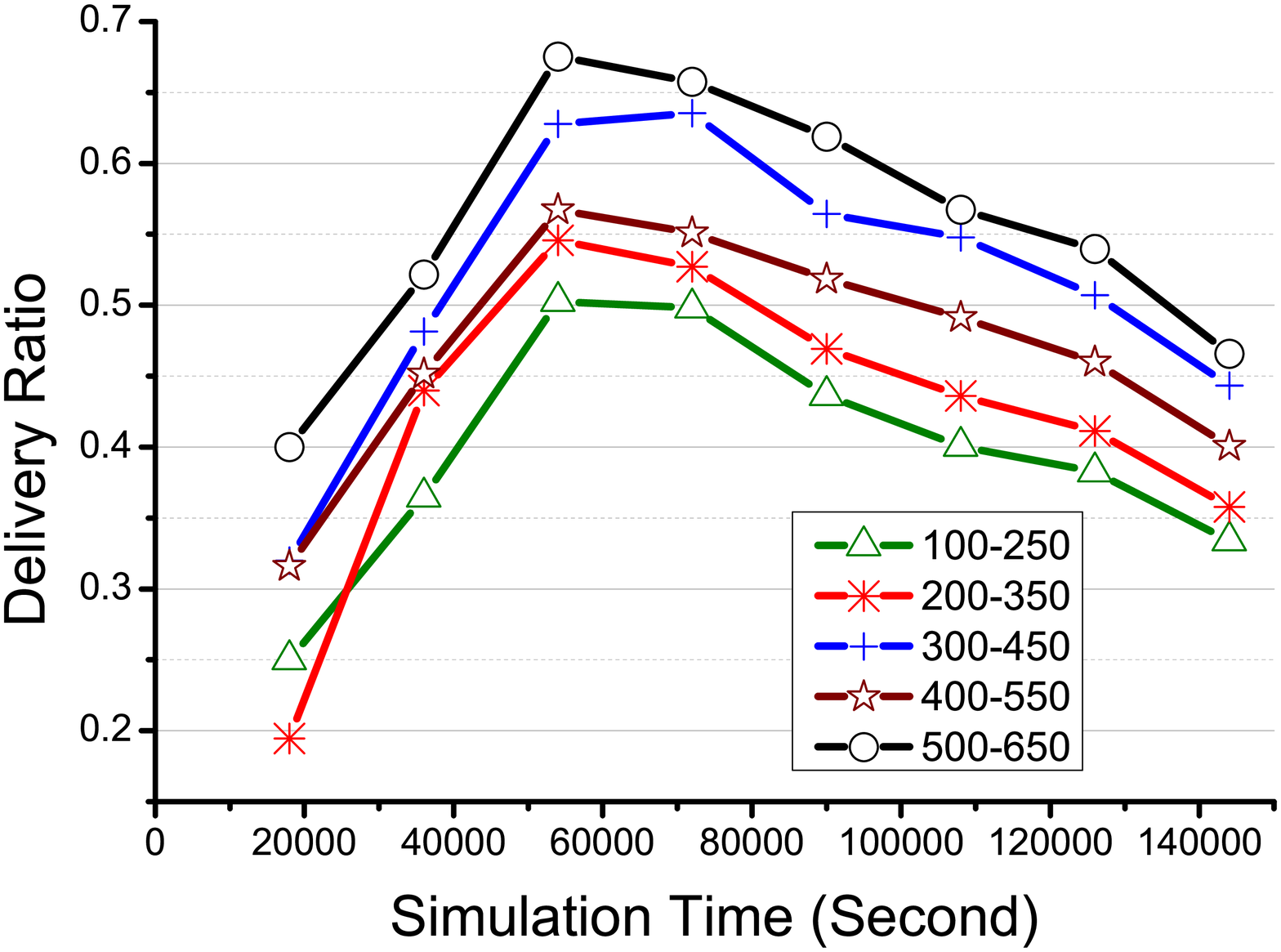}}
\subfigure[Overhead]{
\label{interval-info-overhead}
\includegraphics[width=0.23\textwidth]{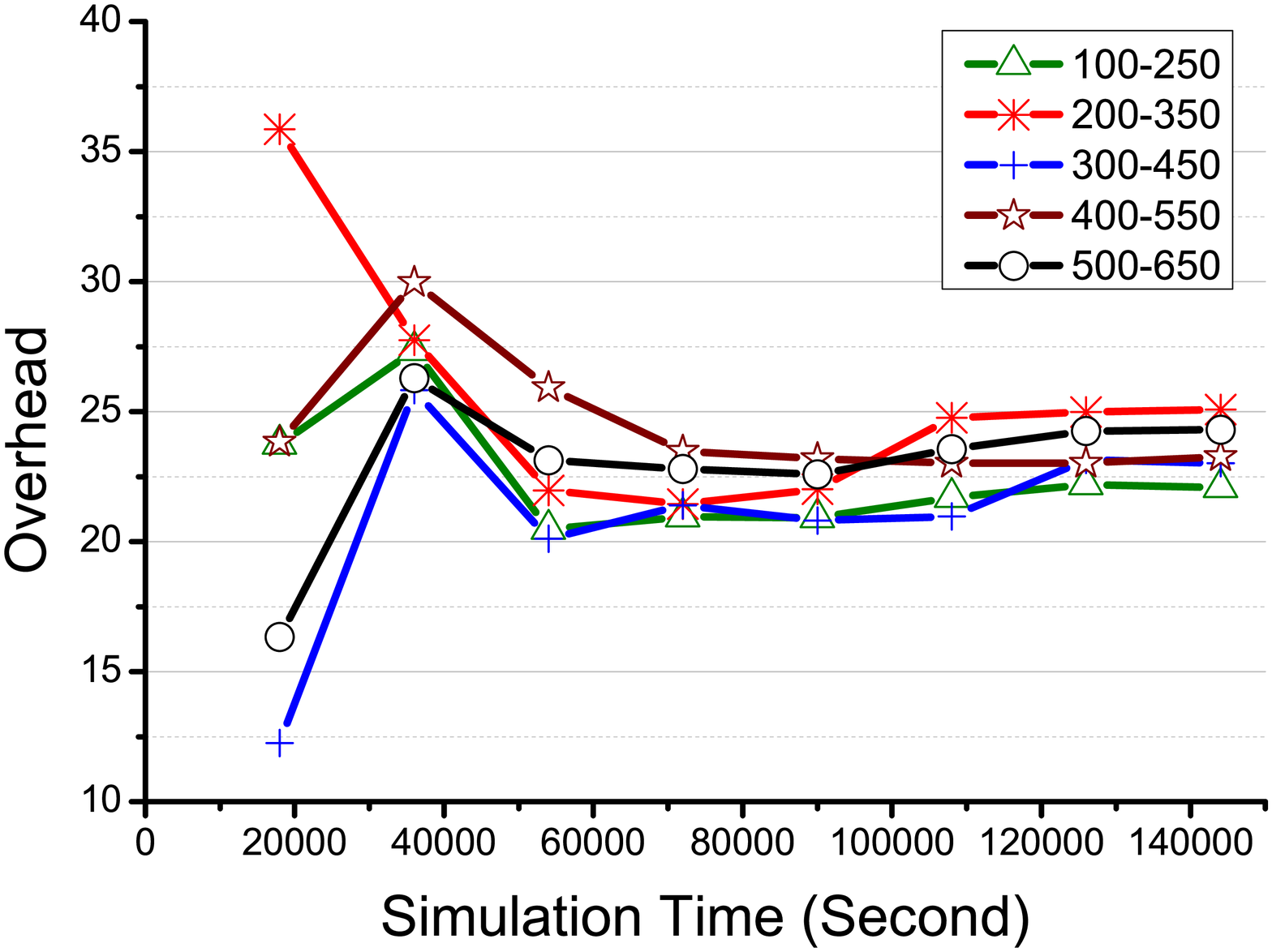}}
\subfigure[Average Latency]{
\label{interval-info-latency}
\includegraphics[width=0.23\textwidth]{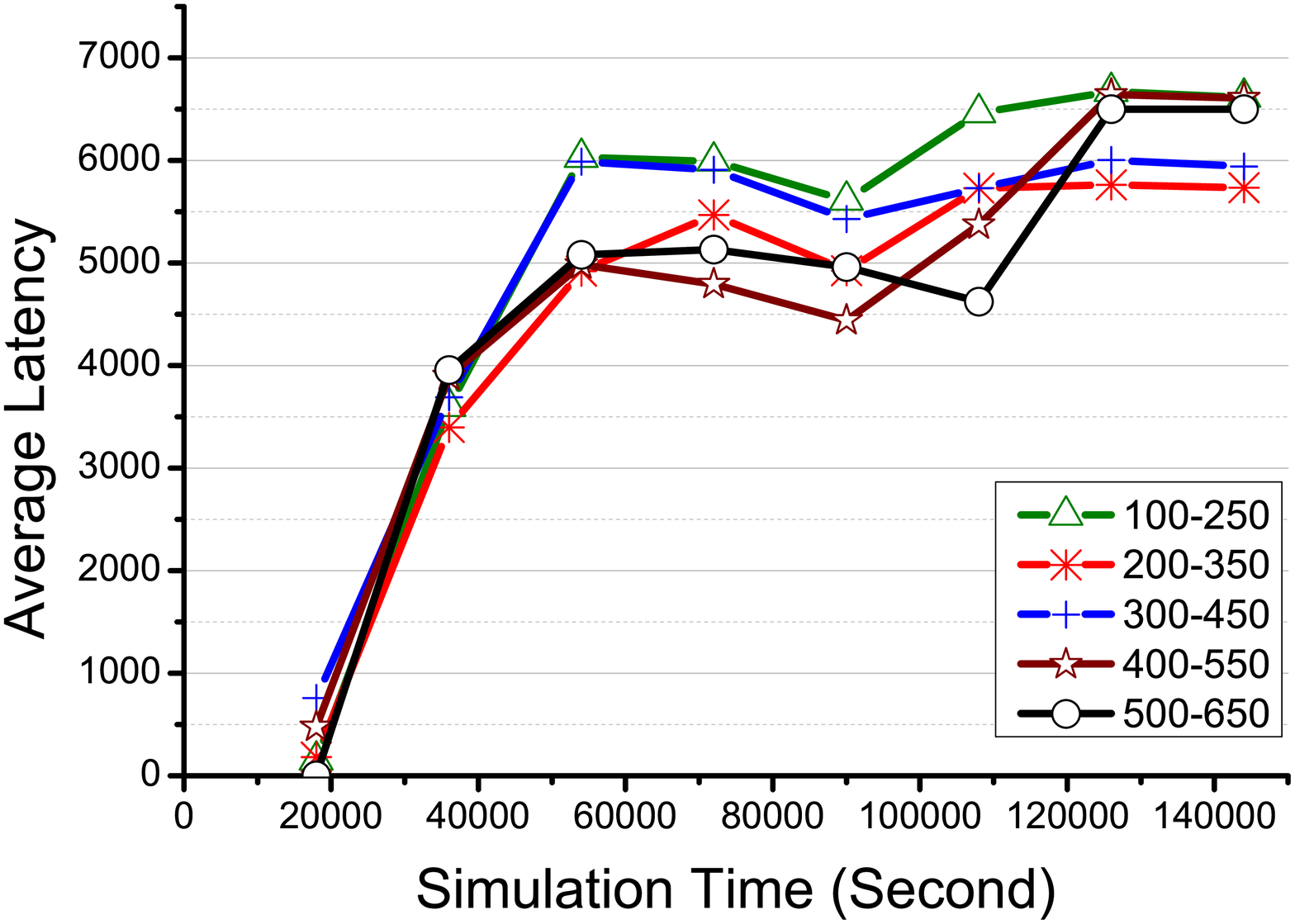}}
\subfigure[Average Hop Count]{
\label{interval-info-hop}
\includegraphics[width=0.23\textwidth]{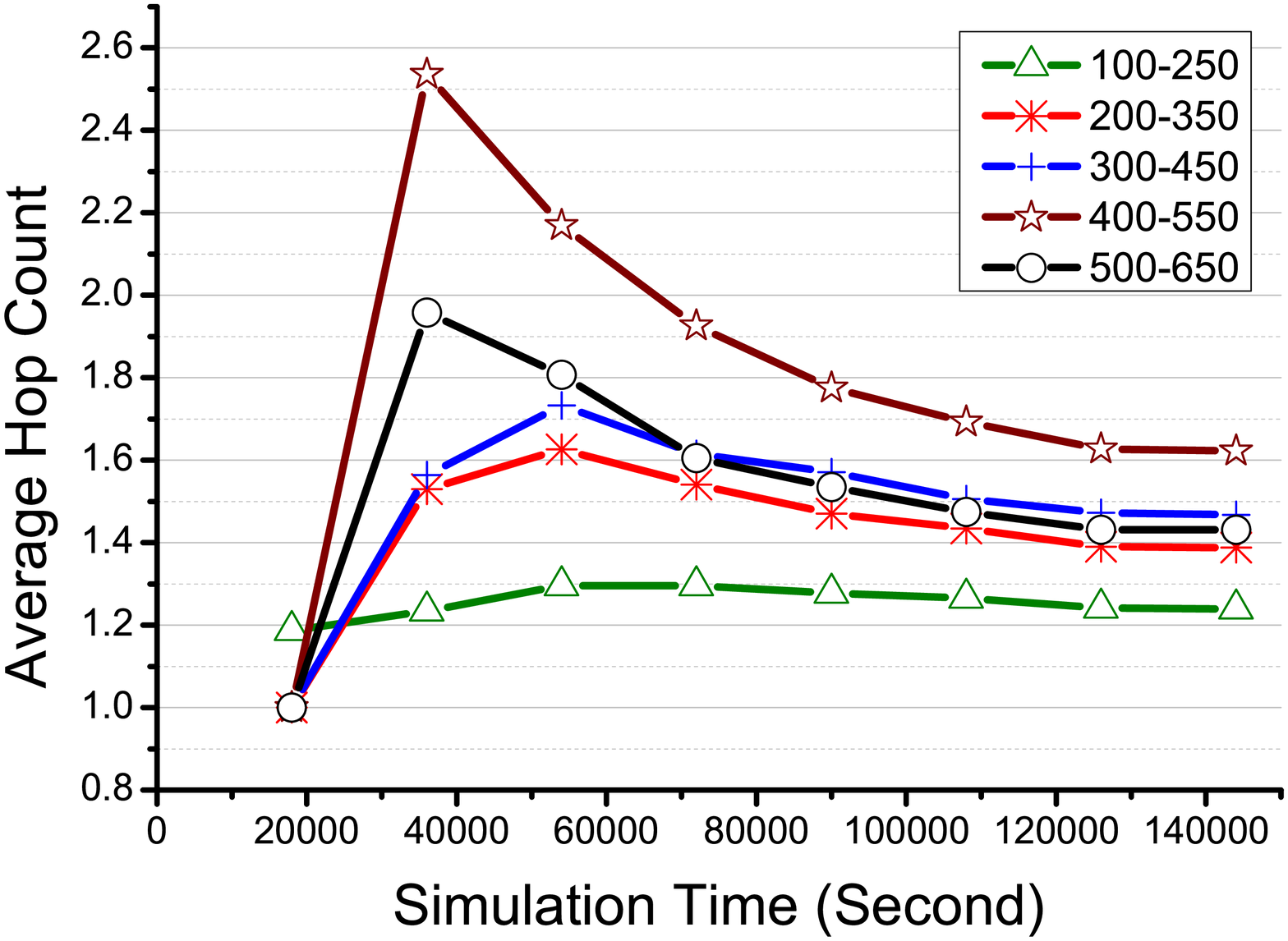}}
\caption{Simulation results for PIS under different message intervals using INFOCOM06 data set.}
\label{interval-info}
\end{figure*}

\subsubsection{The Impacts of Parameters}

There are some parameters which affect the efficiency of the PIS protocol considerably. To this end, we conduct a set of simulations to explore the impact of these parameters.

1. \textit{Parameter $\gamma$}: when two nodes A and B meet each other, their similarities with destination node D with respect to the three social dimensions are calculated. The limited number of messages copies are generated in order to expand the forwarding opportunities. In PIS, the number of message copies is controlled by $nofCopy$, while the transmission range of the message copies is controlled by parameter $\gamma$. If $\gamma$ is set to a small value, the message copies are disseminated even if the similarity gap of nodes A and B is small. As a result, the dissemination of message copies will be finished in a narrow range quickly. Increasing the value of $\gamma$, the transmission range of message copies is enlarged. Due to the limited number of message copies, the bigger $\gamma$ may restrain the copies dissemination which also decreases the contact opportunities. Thus, an appropriate value for $\gamma$ should be selected. We note that in different data sets, the influence of this value can be different according to the nodes' contact frequency.

We compare the performance of the protocols with different $\gamma$ values on the data sets. In SIGCOMM09, $\gamma$ is assigned as 0.2, 0.4, 0.6, and 0.8. While in INFOCOM06, the value of $\gamma$ is set 0.1, 0.2, 0.4, and 0.6. Figs. \ref{gamma-sig} and \ref{gamma-info} show the comparison results, respectively.

In Fig. \ref{gamma-sig}, as the value of $\gamma$ increases, the performance of PIS is improved using SIGCOMM09 data set. However, as shown in Fig. \ref{gamma-info}, PIS achieves the highest performance in INFOCOM06 when $\gamma$ is set to 0.1. With the increasing of value $\gamma$, the performance of PIS declines. Thus, we choose $\gamma$ 0.8 in SIGCOMM09 and 0.1 in INFOCOM06.

\begin{figure*}[!t]
\centering
\subfigure[Delivery Ratio]{
\label{i-sig-delivery}
\includegraphics[width=0.23\textwidth]{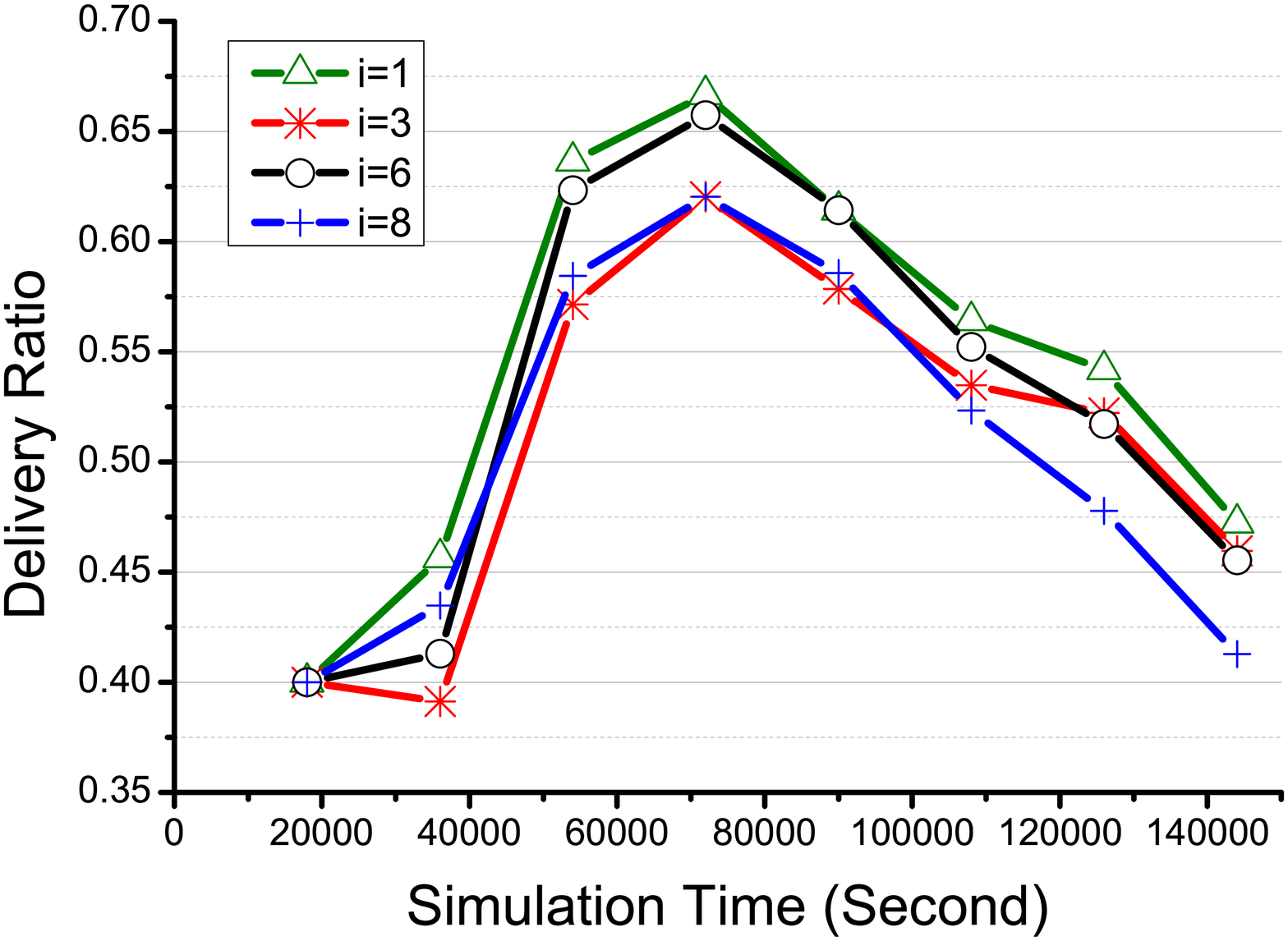}}
\subfigure[Overhead]{
\label{i-sig-overhead}
\includegraphics[width=0.23\textwidth]{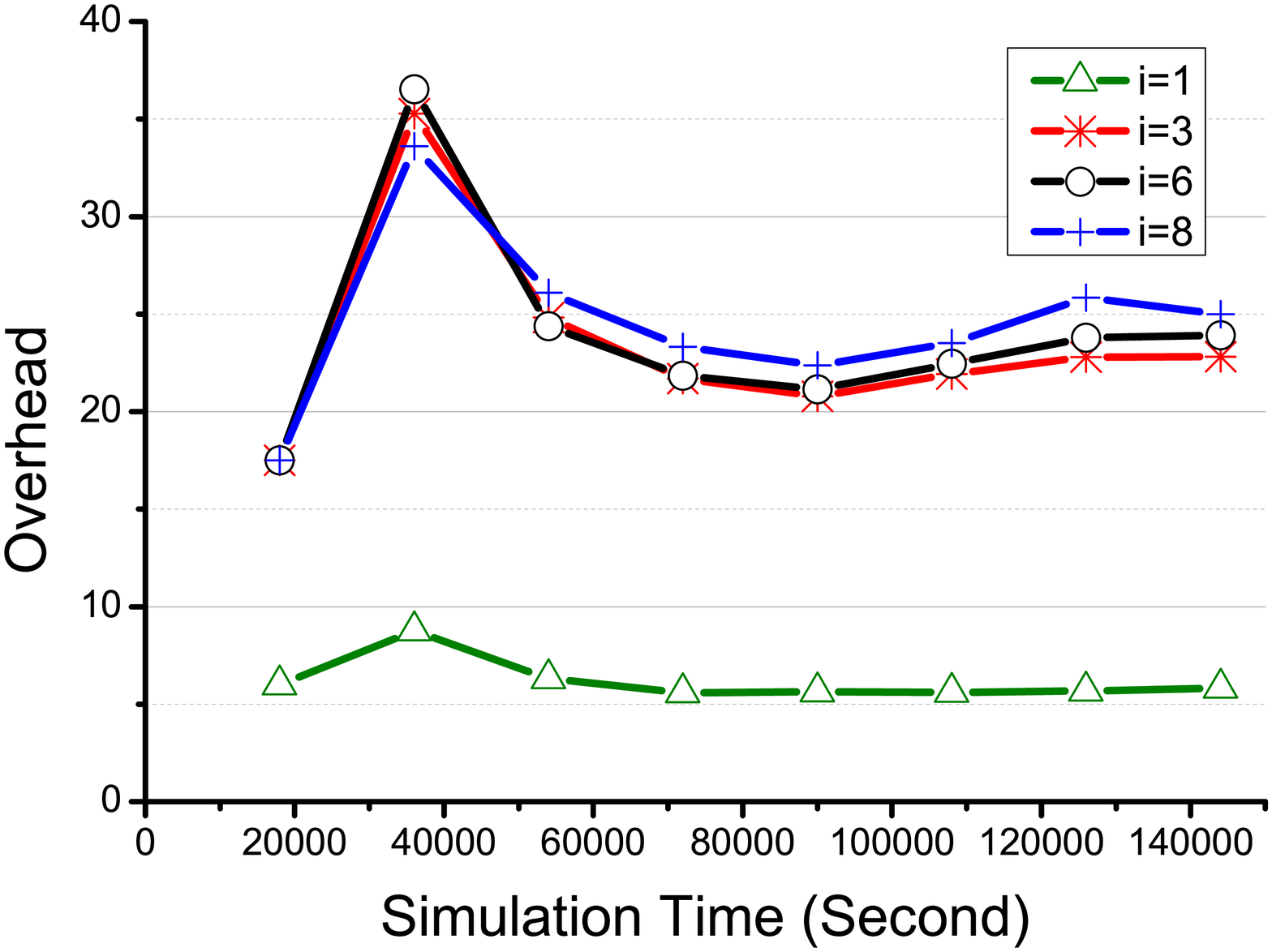}}
\subfigure[Average Latency]{
\label{interval-latency}
\includegraphics[width=0.23\textwidth]{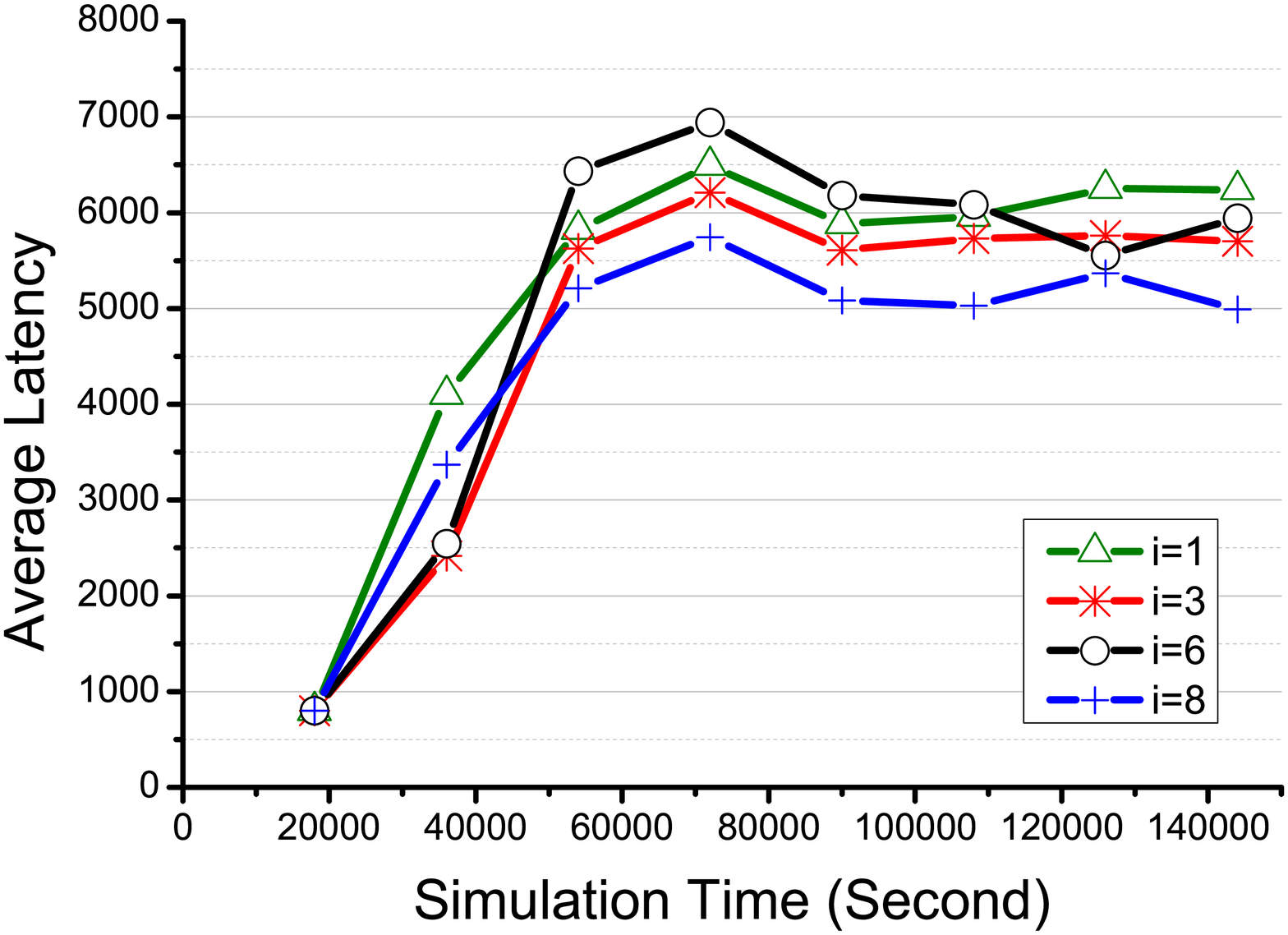}}
\subfigure[Average Hop Count]{
\label{interval-hop}
\includegraphics[width=0.23\textwidth]{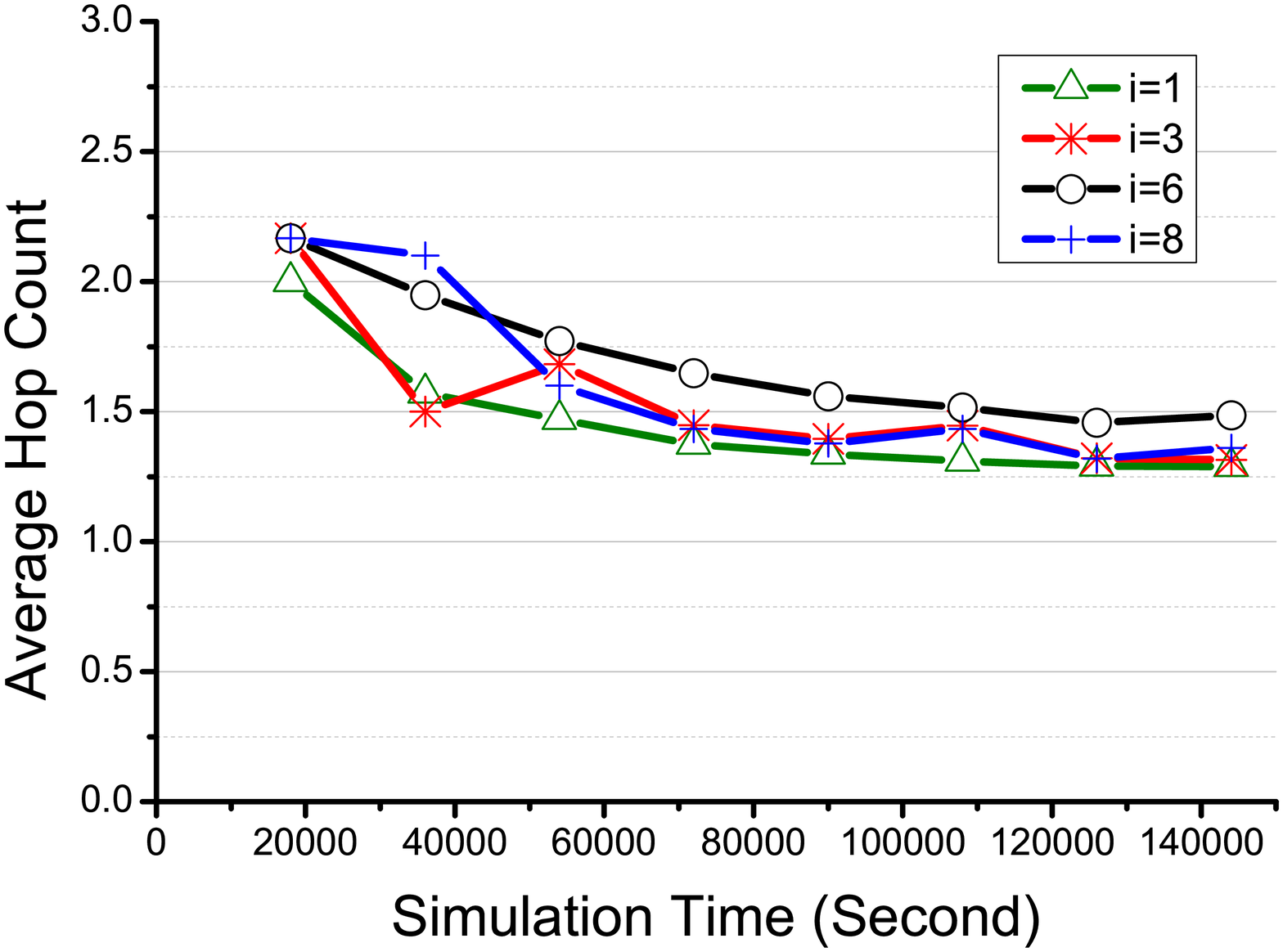}}
\caption{Simulation results for PIS under different time slot parameter \textit{i} using SIGCOMM09 data set.}
\label{i_sig}
\end{figure*}

\begin{figure*}[!t]
\centering
\subfigure[Delivery Ratio]{
\label{i-info-delivery}
\includegraphics[width=0.23\textwidth]{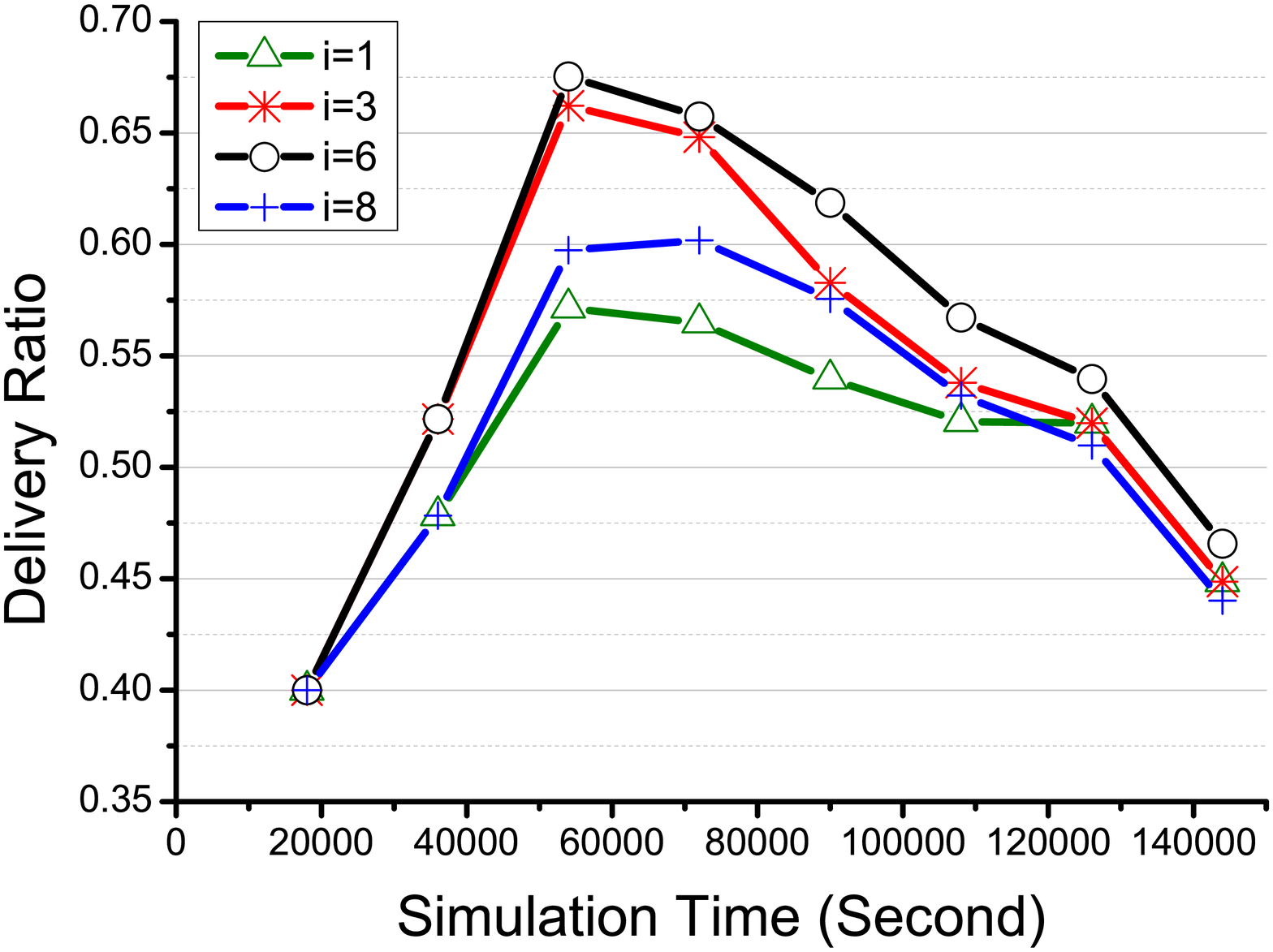}}
\subfigure[Overhead]{
\label{i-info-overhead}
\includegraphics[width=0.23\textwidth]{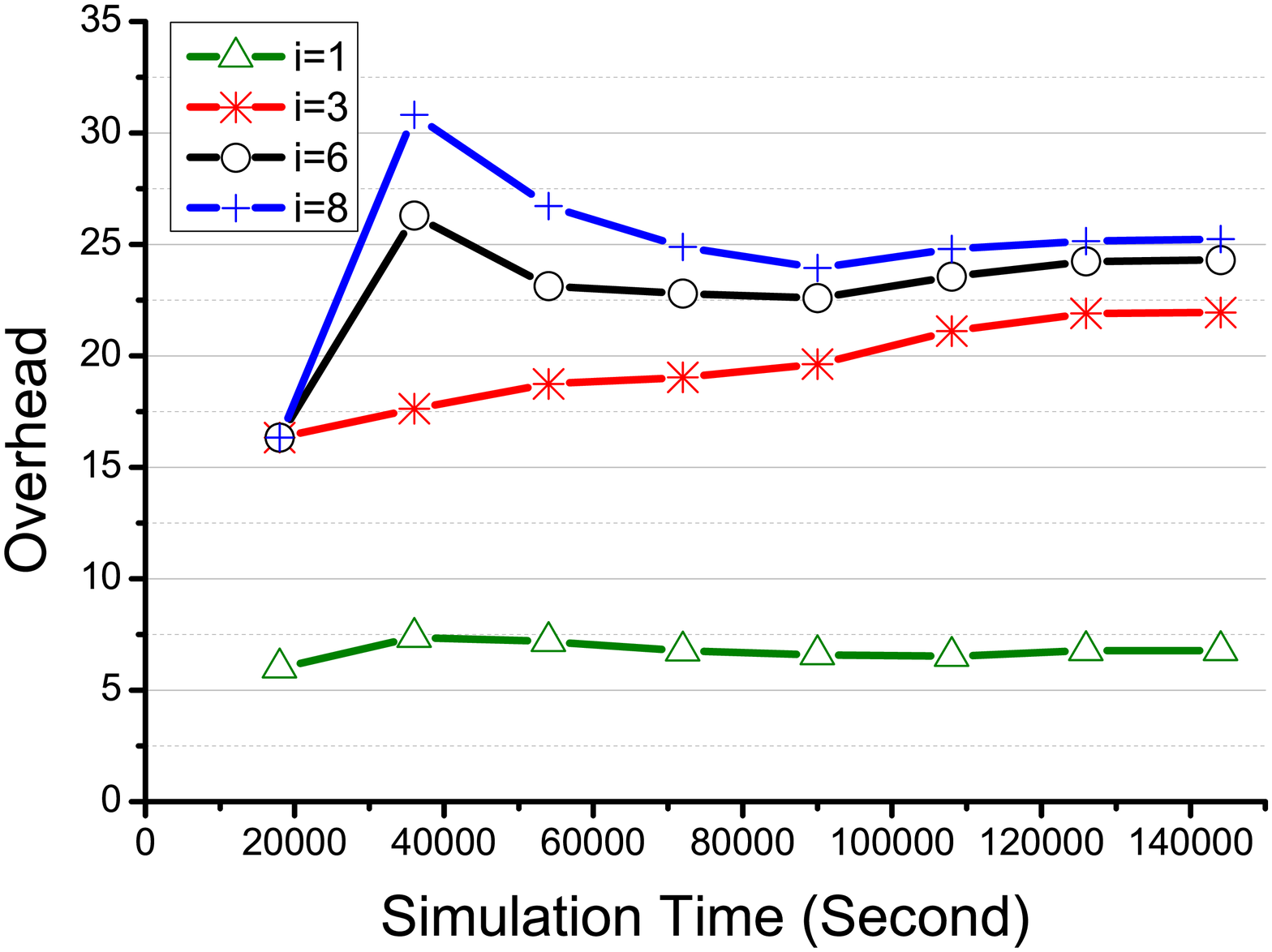}}
\subfigure[Average Latency]{
\label{i-info-latency}
\includegraphics[width=0.23\textwidth]{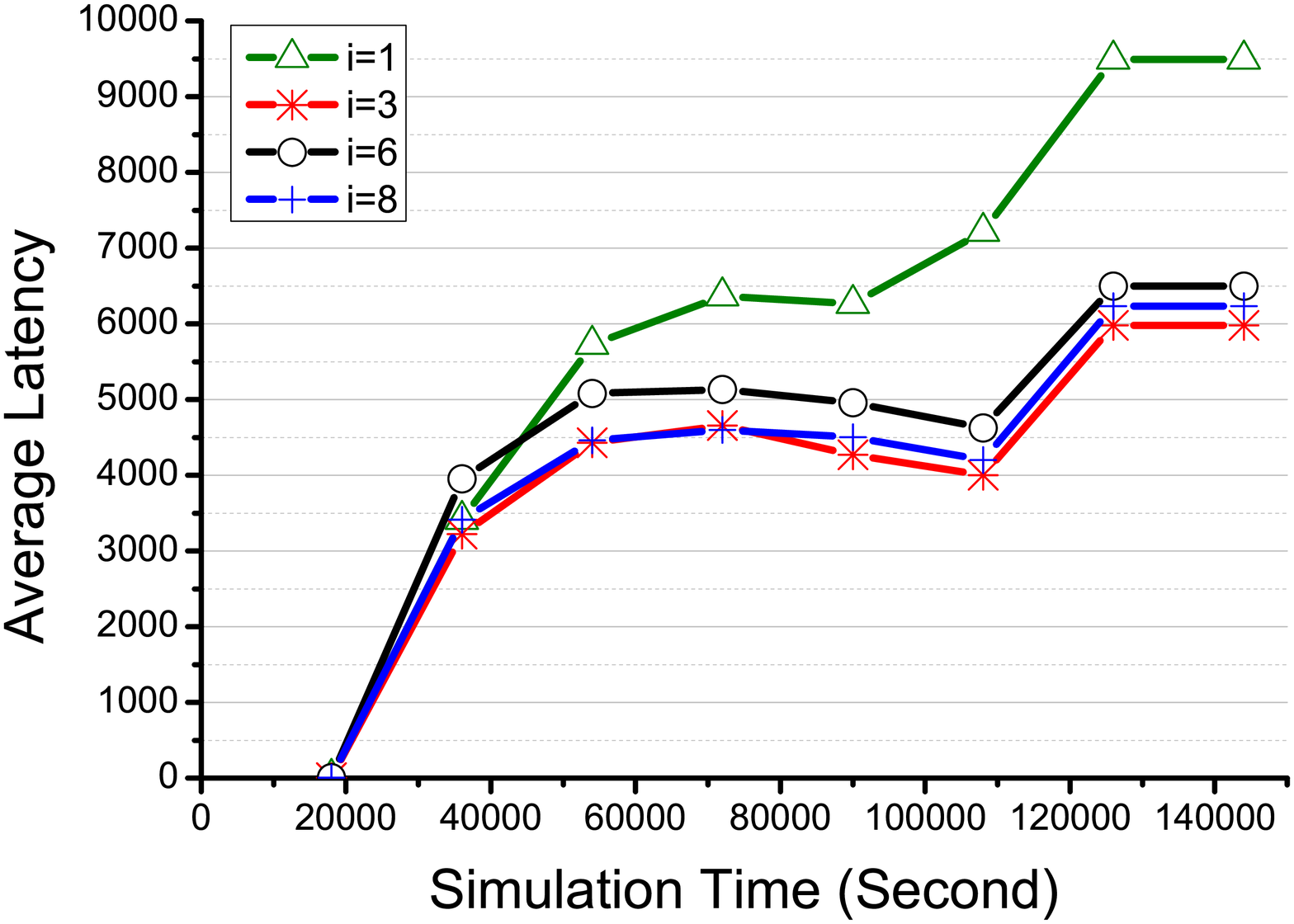}}
\subfigure[Average Hop Count]{
\label{i-info-hop}
\includegraphics[width=0.23\textwidth]{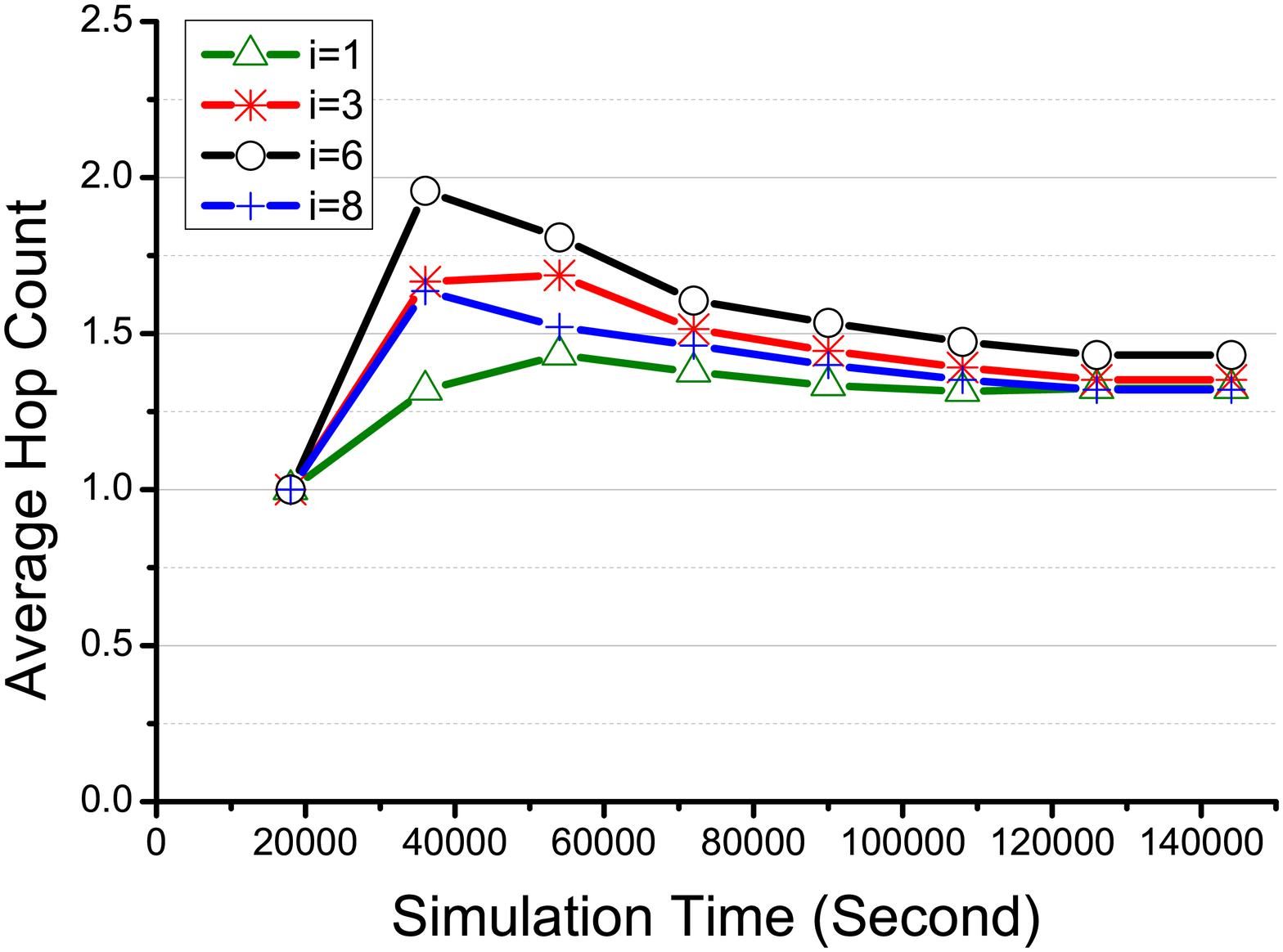}}
\caption{Simulation results for PIS under different time slot parameter \textit{i} using INFOCOM06 data set.}
\label{i_info}
\end{figure*}

2. \textit{Message event interval}: Figs. \ref{interval-sig} and \ref{interval-info} show the influence of message event interval on the performance metrics. Five values are set for message event interval which are 100 $\scriptsize{\sim}$ 250, 200 $\scriptsize{\sim}$ 350, 300 $\scriptsize{\sim}$ 450, 400 $\scriptsize{\sim}$ 550, and 500 $\scriptsize{\sim}$ 650. From this figure, as the number of messages increases, the delivery ratio increases and the average latency prolongs. Moreover, the time interval 500 $\scriptsize{\sim}$ 650 gets the relative higher efficiency due to its lower overhead ratio.

3. \textit{Time slot parameter $i$}: the similarity of two encounter nodes with respect to the three features is the basic forwarding principle in PIS protocol. PIS adopts the time slot mechanism to manage the different time slots. In the simulations, the time slot is set to 1 hour. Time slot parameter $i$ identifies the number of time slots participated in a similarity comparison. In order to analyze the impact of parameter $i$ on PIS, we assign four values to $i$ which are 1, 3, 6, 8. Figs. \ref{i_sig} and \ref{i_info} show the simulation results. As it can be seen from Fig. \ref{i_sig}, $i=6$ and $i=1$ obtain higher performance with the SIGCOMM09 data set. In Fig. \ref{i_info}, PIS achieves the highest efficiency when $i=6$ with the highest delivery ratio, lowest overhead, shortest average latency, and fewer average hop count with the INFOCOM06 data set.

\section{Conclusion}

The social characteristics and behaviors of mobile users have been extensively utilized in the literature to improve the performance of routing protocols in SAN paradigm. In this paper, we proposed a multi-dimensional routing protocol, called PIS, which combines three social features of mobile users with their time regularity in order to design a stable and adaptive forwarding scheme. Our simulation experiments using two real-world mobility data sets have shown that PIS outperforms other benchmark routing protocols (e.g., SimBet, PROPHET, and Epidemic) in terms of data delivery ratio, network overhead, and latency. As part of our future work, we plan to explore the performance of PIS protocol in large-scale networks.

\section*{Acknowledgments}
The authors would like to thank the anonymous reviewers for constructive comments and suggestions which helped improved the quality of the manuscript significantly. This work is partially supported by the Fundamental Research Funds for the Central Universities (DUT15YQ112), the National Natural Science Foundation of China (61572106), and the US National Science Foundation grants CNS-1355505 and CCF-1539318.

\ifCLASSOPTIONcaptionsoff
  \newpage
\fi

\bibliographystyle{IEEEtran}
\bibliography{refs}

\end{document}